\documentclass{iopart}
\def\nb{\nabla}
\def\P{{\it Proof :} \hspace{3mm}} 
\def\N{\hfill \raisebox{1mm}{\framebox{\rule{0mm}{1mm}}}} 

\usepackage{iopams,mathrsfs}  
\eqnobysec
\newtheorem{defi}{Definition}[section]
\newtheorem{theo}{Theorem}[section]
\newtheorem{coro}{Corollary}[section]
\newtheorem{prop}{Proposition}[section]

\newtheorem{Remark}{Remark}[section]
\usepackage[bookmarks,colorlinks=false]{hyperref}
\begin{document}
\title[Dynamical laws of superenergy in General Relativity]
{Dynamical laws of superenergy in General Relativity}

\author{Alfonso Garc\'{\i}a-Parrado G\'omez-Lobo}

\address{Matematiska institutionen, Link\"opings universitet SE-58183 Link\"oping, Sweden}
\ead{algar@mai.liu.se}
\begin{abstract}
The Bel and Bel-Robinson tensors were introduced nearly fifty years ago in an attempt 
to generalize to gravitation the energy-momentum tensor of electromagnetism. 
This generalization was successful from the mathematical point of view because
these tensors share mathematical properties which are remarkably similar to
those of the energy-momentum tensor of electromagnetism. However, 
the physical role of these tensors in General Relativity
has remained obscure and no interpretation has achieved wide acceptance. In principle, 
they cannot represent {\em energy} and the term {\em superenergy} has been
coined for the hypothetical physical magnitude lying behind them.  
In this work we try to shed light on the true physical meaning of {\em superenergy} 
by following the same procedure which enables us to give an interpretation 
of the electromagnetic energy. This procedure consists in 
performing an orthogonal splitting of the Bel and Bel-Robinson tensors 
and analysing the different parts  resulting
from the splitting. In the electromagnetic case such splitting gives rise to 
the electromagnetic {\em energy density}, the Poynting vector and the
electromagnetic stress tensor, each of them having a precise physical interpretation
which is deduced from the {\em dynamical laws} of electromagnetism (Poynting
theorem). The full orthogonal splitting of the Bel and Bel-Robinson tensors 
is more complex but, as expected, similarities with 
electromagnetism are present.  Also the covariant divergence of the Bel tensor is 
analogous to the covariant divergence of the electromagnetic energy-momentum tensor
and the orthogonal splitting of the former is found. The ensuing {\em equations} 
are to the superenergy what the Poynting theorem is 
to electromagnetism.  Some consequences of these {\em dynamical laws of superenergy} 
are explored, among them the possibility of defining  
{\em superenergy radiative states} for the gravitational field.

\end{abstract}

%Uncomment for PACS numbers title message
% Keywords required only for MST, PB, PMB, PM, JOA, JOB? 
%\vspace{2pc}
%\noindent{\it Keywords}: Article preparation, IOP journals
% Uncomment for Submitted to journal title message
%\submitto{\CQG}
% Comment out if separate title page not required
%\maketitle

\section{Introduction}
General Relativity is, in some aspects, a peculiar theory. In it 
the spacetime itself is part of the degrees of freedom and this fact brings
to General Relativity some complications not present in other theories
where the fields are set in a fixed spacetime background. One of these
complications is the impossibility of defining a local invariant 
concept of {\em gravitational energy density}. 
The accepted argument to sustain this
assertion relies on the equivalence principle.  
The consequence of this is that any geometric object 
representing gravitational ``energy-momentum'' 
can always be set to zero in a suitable coordinate system or
frame and this property cannot be fulfilled by a tensor. Only 
a {\em pseudo-tensor} can accomplish this task but gravitational
energy-momentum pseudo-tensors are not unequivocally defined because,
by the very nature of a pseudo-tensor, they are always tied to a 
given frame or coordinate system. The use of a pseudo-tensor makes it very
difficult to address problems such as the calculation
of the gravitational energy radiated by a source. 

Different approaches to the ``gravitational energy problem'' 
in General Relativity
have been provided along the years and no general formalism has 
emerged (although formalisms tailored for particular important 
cases do exist). One of these approaches 
seeks to enhance the formal similarities between electromagnetism 
and gravitation in order to find a replacement for the 
missing ``gravitational energy-momentum tensor''.
The idea is to take the electromagnetic energy-momentum
tensor and translate it into a gravitational counterpart by somehow replacing 
the Faraday tensor with the Riemann tensor in the expression giving the 
energy-momentum tensor for electromagnetism. This translation is by no
means straightforward due to the different nature of 
the Riemann and Faraday tensors but it can certainly be 
accomplished. The result of this translation is 
a four index tensor quadratic in the Riemann tensor which
was first found by Bel \cite{BEL2}. The Bel tensor
has mathematical properties which are remarkably similar to 
the electromagnetic energy-momentum tensor 
(see theorem \ref{bel-properties} for a summary). An important particular
case arises if we replace in the definition of the Bel tensor
the Riemann with the Weyl tensor to give the  
Bel-Robinson tensor \cite{BEL3}.  

From the above considerations it is clear that the Bel tensor
will represent a quantity which is different from energy.
This new quantity was called ``superenergy'' by Bel and its status
in General Relativity has been subject to much debate and no widely 
accepted conclusions have been reached. A simple dimensional 
analysis shows that in geometrized units the physical dimension of superenergy
is $L^{-4}$ where $L$ represents length. 
Another important property is the tensorial character of 
superenergy. This means that if we work with 
{\em gravitational superenergy} instead of {\em gravitational energy} 
we can avoid all the technical complications
arising when one works with pseudotensors.
One of the main goals of this paper is to show what the consequences are 
of considering superenergy as a measurable physical quantity on its own.
This means that we are not concerned in this work with the 
possible relationship between superenergy and other
quantities with
dimensions of energy. 

In order to carry out our program we need to find the orthogonal 
splitting with respect to an observer of the Bel tensor (so we will
be able to explain what the observer obtains when measuring superenergy) 
and we need to find the variation of the different parts of the
orthogonal splitting along the observer's path. The outcome
of this last part is a set of equations which we call {\em 
the dynamical laws of superenergy} and they are the most 
important result of this paper.
    
We may examine at this point what the above procedure yields in the case of 
electromagnetism. In this case we are working with a quantity
with dimensions of energy instead of dimensions of superenergy but this is now 
of no relevance. The different parts 
resulting from the orthogonal splitting of the electromagnetic
energy-momentum tensor are the electromagnetic energy density, the Poynting
vector and the electromagnetic stress-tensor. The utility of each of these 
parts is explained in basic electrodynamics textbooks. The 
{\em dynamical laws of electromagnetic energy} are contained in the Poynting
theorem and it is through this theorem that the electromagnetic energy density
and the Poynting vector gain their full physical 
meaning as measurable quantities. The Poynting theorem
is nothing less than the orthogonal splitting of the covariant divergence
of the electromagnetic energy-momentum tensor. The parts of this splitting are
the variation of the electromagnetic energy density and the Poynting vector 
along the observer's path. The Poynting theorem enables us to draw conclusions
as important as the characterization of radiative electromagnetic fields or
the expression for the total force acting on an electromagnetic system. 

In General Relativity we may consider the expression for the covariant 
divergence of the Bel tensor as the gravitational counterpart of the covariant
divergence of the energy momentum tensor of electromagnetism. Therefore if
we perform the orthogonal splitting of the former we will obtain a set of 
equations which can be regarded as the counterpart of the Poynting theorem.
As mentioned before these equations are the dynamical laws of superenergy
and they are far more complex than electromagnetism's Poynting theorem. 
However, we can still follow the same 
procedure as in electromagnetism to draw some conclusions and, for example,
we can decide in a covariant way
when a gravitational system is radiating superenergy 
(intrinsic superenergy radiative state). This was already attempted by 
Bel in the late fifties but since the full set of dynamical laws of
superenergy was not available, Bel's result does not apply to cases that are
sufficiently general.

The paper is organized as follows: in section \ref{essentials}
we review the notation and the essential concepts of orthogonal 
splittings. In section \ref{em-example} we find the orthogonal splitting
of the covariant divergence of the electromagnetic energy-momentum tensor
in a general spacetime (theorem \ref{poynting-theorem}). 
This is the complete version of the classical Poynting theorem and 
some of its consequences are discussed. In section 
\ref{gravitational-equations} we present the Bel and Bel-Robinson tensors
and their essential mathematical properties are summarized in 
theorem \ref{bel-properties}. Section \ref{br-splitting} contains the 
orthogonal splitting of the Bel-Robinson tensor and we study 
the basic mathematical properties of the different parts 
of the orthogonal splitting. Since these
parts are expressed in terms of the electric and magnetic parts 
of the Weyl tensor,
we can obtain particular {\em canonical forms} valid for some Petrov types
(subsection \ref{petrov-forms}). Section \ref{bel-splitting} is devoted
to the orthogonal splitting of the Bel tensor and section 
\ref{superenergy-evolution} contains the main result of this paper which is 
theorem \ref{superenergy-law}. This theorem spells out the different
parts of the orthogonal decomposition of the covariant divergence of the 
Bel tensor (see equation (\ref{mattercurrent-conservation})) which as
explained above are the dynamical laws of superenergy. In section
\ref{superenergy_balance} we study the radiation of superenergy from
a general point of view. To that end the definition of an {\em intrinsic
superenergy radiative state} is put forward (definition \ref{radiation-state}).

The main results of this paper rely on heavy tensor 
calculations which can only be 
carried out with the aid of a computer algebra system. All the calculations of 
this paper have been undertaken with the computer program {\em xAct}
\cite{JMM}. {\em xAct} is a suite of MATHEMATICA packages 
devised to perform calculations in General Relativity and Differential Geometry.
Among the many features of the {\em xAct} system we stress its ability to 
canonicalize tensor expressions by means 
of powerful algorithms based on permutation
group theory (package {\em xPerm}), the excellent implementation of tensor
calculus (package {\em xTensor}) and the possibility of working with frames 
and tensor components (package {\em xCoba}). In appendix A we provide 
further details about how {\em xAct} has been used in this paper. Currently,
no other computer algebra system, either free or commercial, 
has the capabilities to perform the calculations needed in this paper.

\section{The orthogonal splitting}
\label{essentials}
We start by introducing the basic notation and conventions which 
will be adopted in this paper. 
We shall work in a four dimensional smooth Lorentzian manifold $V$ which we
will call {\em spacetime}.
The abstract index notation is followed throughout to denote tensors 
on $V$ with Latin lowercase letters reserved for the abstract indices.  
We use bold typeface for component indices.
Round (square) brackets enclosing indices denote index symmetrization 
(antisymmetrization).
Unless otherwise stated all tensors are assumed smooth and defined globally on $V$.
The metric tensor  is ${\rm g}_{ab}$ and our signature convention is $(-,+,+,+)$. 
This metric is used to raise and lower indices in the usual way. 
Associated with the metric is the {\em volume element} which we denote by
$\eta_{abcd}$.
The Levi-Civita connection compatible 
with ${\rm g}_{ab}$ is the only affine connection $\nabla_a$ 
satisfying $\nb_a{\rm g}_{bc}=0$ 
and our convention for the curvature tensor of this 
connection is fixed by the Ricci identity  
$$
\nb_{a}\nb_{b}X^c-\nb_b\nb_aX^c=X^dR_{bad}^{\ \ \ c}.
$$
The Ricci tensor and the scalar curvature are $R_{bd}\equiv R_{bad}^{\ \ \ a}$ 
and $R\equiv R^a_{\ a}$ 
respectively. From these, the Einstein tensor is defined by the familiar 
formula $G_{ab}\equiv R_{ab}-R{\rm g}_{ab}/2$. 
The Lie derivative with respect to
any vector field $X^a$ is the differential operator $\pounds_X$. 
Geometrized units with $8\pi G=c=1$ are used unless otherwise stated. 
The end of a proof is 
marked with \raisebox{1mm}{\framebox{\rule{0mm}{1mm}}}.

Specially important for us are unit timelike vector fields. For any such vector field,
the family of its integral curves defines a {\em timelike congruence} or 
{\em observer set}. This unit timelike vector field enables us to perform an 
{\em orthogonal splitting} (also called {\em 3+1 decomposition}) 
of any tensor on $V$. The orthogonal splitting lies at the basis 
of many studies and formalisms in General
Relativity and has been extensively studied in the literature but since 
it will be used in this work many times
we now review its essentials (good accounts can be found in 
\cite{ELLIS,MAR-BAS}).
 Let $n^a$ be any vector field with $n_an^a=-1$
and define the {\em spatial metric} $h_{ab}$ by
\begin{equation}
h_{ab}\equiv g_{ab}+n_an_b,\ h_{ab}h^{b}_{\ c}=h_{ac},\ h^a_{\ a}=3.
\label{spatial-metric}
\end{equation}
The tensor $h_{ab}$ has the properties of an orthogonal projector. 
We shall call a covariant tensor $T_{a_1\dots a_m}$ {\em spatial} with respect to $h_{ab}$ 
if it is invariant under $h^a_{\ b}$ i.e. if
$$
h^{a_1}_{\ b_1}\cdots h^{a_m}_{\ b_m}T_{a_1\cdots a_m}=T_{b_1\cdots b_m}, 
$$
with the obvious generalization for any mixed tensor. 
This property implies that the inner contraction of $n^a$ with 
$T_{a_1\dots a_m}$ (taken on any index) vanishes.
We introduce next the orthogonal projection operator defined by
\begin{equation}
P_h(L_{a_1\dots a_m})\equiv h_{\ a_1}^{s_1}\cdots h_{\ a_r}^{s_m}
L_{s_1\dots s_m},
\label{p-operator}
\end{equation}
where $L_{a_1\dots a_m}$ is an arbitrary tensor. Clearly 
$P_h(L_{a_1\dots a_m})$ is a spatial tensor. Another definition which
we need is the {\em generalized inner contraction} of the tensor $L_{a_1\dots a_m}$ 
with the unit normal which is given by 
$$
n^{J}(L_{a_1\dots a_m})\equiv n^{s_1}\cdots n^{s_{\#J}}L_{\dots s_1\dots s_2\dots s_{\#J\dots}}.  
$$ 
Here $J$ is an ordered subset of the set of abstract indices $\{a_1\dots a_m\}$
and the dummies $\{s_1\dots s_{\#J}\}$ are placed in those slots of $L$
indicated by $J$. Therefore $n^J(L_{a_1\dots a_m})$ has $m-\#J$ free indices given by the
complement of $J$ with respect to $\{a_1\dots a_m\}$.
Using the orthogonal projection operator and the 
generalized inner contraction we find that any tensor
$L_{a_1\dots a_m}$ can be written in the following way 
\begin{equation}
\fl L_{a_1\dots a_m}=\sum_{J\in{\mathscr P}(\{a_1\dots a_m\})}
(-1)^{\# J}n_JP_h(n^J(L_{a_1\dots a_m})),\ 
\label{general-ot}
\end{equation}
where ${\mathscr P}(\{a_1\dots a_m\})$ is the power set of $\{a_1\dots a_m\}$ and
$$
n_J\equiv n_{a_q}\cdots n_{a_p},\ J=\{a_q,\dots,a_p\}
\in{\mathscr P}(\{a_1\dots a_m\}).
$$
The right hand side of (\ref{general-ot}) is called the 
{\em orthogonal splitting} of $L_{a_1\dots a_m}$ with respect to
the unit normal $n^a$ (we will just speak of orthogonal splitting of a tensor 
if the unit normal is understood). 
The orthogonal splitting given by (\ref{general-ot}) is 
unique and the set of spatial tensors 
$\{P_h(n^J(L_{a_1\dots a_m}))\}$ contains all the information about $L_{a_1\dots a_m}$.
Equation (\ref{general-ot}) is just the traditional calculation of the orthogonal
splitting of a tensor written in a short form. It is possible to study the
orthogonal splitting of a general tensor in an alternative way if we regard it
as a {\em $r$-fold form} (see \cite{SUPERENERGY,EHDECOMPOSITION} for a precise explanation of
this).

A trivial example of orthogonal splitting is that of 
the metric tensor itself which is obtained from the 
first expression in (\ref{spatial-metric}). 
Another important example of orthogonal splitting which is easily deduced from
(\ref{general-ot}) is 
$$
\eta_{abcd}=-n_a\varepsilon_{bcd}+n_b\varepsilon_{acd}
-n_c\varepsilon_{abd}+n_d\varepsilon_{abc},
$$
where $\varepsilon_{abc}$ is the {\em spatial volume element} 
and is defined by
$$
\varepsilon_{abc}\equiv n^d\eta_{dabc}.
$$

\subsection{Kinematical quantities}
As we explained above, the set of integral curves of $n^a$ 
represents a family of observers. In 
physical applications it is important to introduce quantities describing 
the {\em relative motion} of each curve of the family and this is the role of
the kinematical quantities. To define them we write down the orthogonal splitting of 
$\nb_an_b$ which is    
\begin{equation}
\nb_an_b=-A_bn_a+\frac{1}{3}\theta h_{ab}+\sigma_{ab}+\omega_{ab}.
\label{decompose-normal}
\end{equation}
The tensor $A_b$ is the acceleration, the scalar $\theta$ is the expansion and
$\sigma_{ab}$, $\omega_{ab}$ are the shear and the rotation respectively.  
From the previous equation it is easy to obtain expressions for the kinematical
quantities in terms of $n^a$ 
\begin{equation}
\fl A^b=n^a\nb_{a}n^b,\ \theta=\nb_an^a,\ 
\omega_{ab}=h_{[a}^{\ d}h_{b]}^{\ c}\nb_dn_c,\ 
\sigma_{ab}=h_{(a}^{\ d}h_{b)}^{\ c}\nb_dn_c-\frac{\theta}{3}h_{ab}.
\end{equation}
Straightforward properties of the kinematical quantities are
\begin{equation}
\sigma_{(ab)}=\sigma_{ab},\ 
\omega_{[ab]}=\omega_{ab},\ 
\sigma^a_{\ a}=0,
\label{k-properties}
\end{equation}
Sometimes the rotation is replaced by the {\em vorticity} which is defined
as follows
\begin{equation}
\omega_a\equiv\frac{1}{2}\varepsilon_{abc}\omega^{bc}\ \Rightarrow
\omega_{ab}=\varepsilon_{abc}\omega^c.
\end{equation}  
Each of the kinematical quantities has a precise interpretation which deals 
with the {\em relative motion} of the observers of the congruence 
(see e.g. \cite{ELLIS,HAWKING-ELLIS} for a
more detailed description of these concepts).
\subsection{Cattaneo operator}
\label{section-cattaneo}
Another very important object, which is needed when working with orthogonal 
splittings is the {\em Cattaneo operator} also known as {\em spatial connection} 
\cite{CATTANEO}. 
If $L_{a_1\dots a_m}$ is a covariant tensor
then we define the linear operator
\begin{equation}
D_aL_{a_1\dots a_m}\equiv 
P_h(\nb_aL_{a_1\dots a_m}),
\label{cattaneo}
\end{equation}
with obvious definitions for contravariant and mixed tensors. The Cattaneo 
operator is not a linear connection on the spacetime manifold 
$V$ because it does not satisfy the Leibnitz
rule unless both factors of the product upon which $D_a$ acts are spatial.
From its definition, it is clear that $D_aT_{b_1\dots b_m}$ is a spatial 
tensor.  
Important additional properties of the Cattaneo operator are
\begin{equation}
\fl D_ah_{bc}=0,\ D_aD_b\varphi-D_bD_a\varphi=2n^c\omega_{ab}\nb_c\varphi=
2\omega_{ab}\pounds_{n}\varphi,\ \varphi\in C^{1}(V).
\end{equation} 

The Cattaneo operator enables us to write in a 
compact form the orthogonal splitting of any
expression involving derivatives. It is specially important in this 
work to find the orthogonal splitting of
the covariant derivative of a spatial tensor.  
To illustrate how this works, let us 
study a particular simple example. 
Consider $\nb_aL_b$, where $L_b$ is an arbitrary
spatial covector ($n^aL_a=0$). In this case formula (\ref{general-ot}) yields
\begin{equation}
\fl\nb_aL_b=D_aL_b-n_aP_h(n^p\nb_pL_b)-n_bP_h(n^p\nb_aL_p)+
n_an_bn^pn^q\nb_pL_q.
\label{first-stage}
\end{equation}
Next we use in this equation the relations
\begin{eqnarray}
n^c\nb_cL_a=\pounds_nL_a-L_c\nb_an^c,\ 
n^p\nb_aL_p=-L_p\nb_an^p,
\end{eqnarray}
and replace in (\ref{first-stage}) the covariant derivatives 
of the unit normal by
the expression given in (\ref{decompose-normal}). After some manipulations
equation (\ref{first-stage}) becomes
\begin{eqnarray}
\fl\nb_aL_b=-A^c L_cn_bn_a +\left(\frac{1}{3} L_b 
   \theta-\pounds_nL_b+L^c (\sigma_{bc}
+\omega_{bc})\right) n_a+D_aL_b+\nonumber\\
+n_b\left(\frac{1}{3}L_a\theta+L^c\left(\sigma_{ac}+\omega_{ac}\right)\right),
\label{l-split}
\end{eqnarray}
which has the form of (\ref{general-ot}) and hence is 
the complete orthogonal splitting of $\nb_aL_b$. Note that 
$\pounds_nL_a$ is a spatial covector if $L_a$ is spatial, due to the 
property $\pounds_nn^a=0$. The procedure followed to obtain 
(\ref{l-split}) is easily generalized for the covariant 
derivative of any spatial tensor (see appendix A for more 
examples). This kind of calculation 
is extensively used in section \ref{superenergy-evolution}. 

\section{Electromagnetism as a working example} 
\label{em-example}
As a preparation for the study which we are going to undertake of 
the gravitational field, we analyse first the case of electromagnetism.
The electromagnetic field is described by an antisymmetric rank-2 tensor $F_{ab}$
(the Faraday or electromagnetic tensor) which satisfies the {\em Maxwell equations}
\begin{equation}
\nb_{[a}F_{bc]}=0,\ \nb_aF^a_{\ b}=j_b,
\label{maxwell}
\end{equation}
 where $j^b$ is {\em the charge current} four vector (here we follow
the Heaviside-Lorentz units system).
A very important object in electromagnetic theory is the energy-momentum
tensor of the electromagnetic field, given by
\begin{equation}
T_a^{\ b}=\frac{1}{2}(F_{ad}F^{bd}+F^*_{ad}F^{*bd})=
F_{ad}F^{bd}-\frac{1}{4}\delta^a_{\ b}F_{cd}F^{cd}.
\label{energy-electromagnetic}
\end{equation}
In this formula $F^*_{ab}$ is the Hodge dual of $F_{ab}$ defined by
$$
F^*_{ab}\equiv\frac{1}{2}\eta_{abcd}F^{cd}.
$$
\begin{theo}
The tensor $T_{ab}$ has the following properties
\begin{enumerate}
\item $T_{(ab)}=T_{ab}$.
\item $T_{ab}$ always satisfies the {\em dominant energy 
condition}, namely, for any pair $u^a$, $v^a$ of causal future-directed
vector fields the inequality $T_{ab}u^av^b\geq 0$ holds. 
\item If the Maxwell equations hold then we have
\begin{equation}
\nb_bT^b_{\ a}=F_a^{\ b}j_b,\ \nb_aj^a=0.
\label{e-mconservation}
\end{equation}
\end{enumerate}
\label{maxwell-properties}
\end{theo}
Roughly speaking, the first equation of (\ref{e-mconservation}) tells us that the 
variation of the electromagnetic energy-momentum equals the work
performed by the charge current and the second equation is the equation of charge
conservation. An adequate understanding of these informal
assertions can be achieved by finding the orthogonal splitting of 
(\ref{e-mconservation}). As an aside remark, we note that the first equation
of (\ref{e-mconservation}) is not in general equivalent to Maxwell equations as is 
sometimes wrongly stated. 
 
To find the orthogonal splitting of (\ref{e-mconservation}) we first
need to find the orthogonal splitting of $F_{ab}$. Define the spatial tensors
\begin{equation}
E_a\equiv F_{ab}n^b,\ B_a\equiv F^*_{ab}n^b
\end{equation}
These are the electric and magnetic parts of 
the Faraday tensor and they characterize it completely. The orthogonal decomposition
of the Faraday tensor in terms of $E_a$ and $B_a$ reads
\begin{equation}
F_{ab}=E_bn_a-E_an_b-B_p\varepsilon_{ab}^{\ \ p}.
\end{equation}
Using this expression, we can find the orthogonal splitting of the energy-momentum
tensor $T_{ab}$ which results in 
\begin{eqnarray}
\fl 
T_{ab}=Un_an_b+2P_{(a}n_{b)}+{\mathcal T}_{ab},\ \nonumber\\
\fl U\equiv\frac{1}{2}(E_aE^a+B_aB^a),\ 
P_a\equiv\varepsilon_{abc}B^bE^c,\ 
{\mathcal T}_{ab}\equiv Uh_{ab}-E_aE_b-B_aB_b.
\label{tem-decomposition}
\end{eqnarray}
Also, the orthogonal splitting of $j^{a}$ is easily found yielding
$$
j^{a}=\rho n^{a}-J^{a},\ \rho\equiv -j^{a}n_{a},\ J^{a}\equiv -h^{ac}j_{c}
$$
where $\rho$ is the charge density and $J^{a}$ is the spatial charge current.
Next we replace the decomposition of $T_{ab}$ and $j^a$ in 
(\ref{e-mconservation}) and 
calculate the orthogonal splitting of the resulting equations. 
To achieve this we need to find the orthogonal splitting of $\nb_aU$, 
$\nb_aP_b$ and $\nb_a{\mathcal T}_{bc}$ 
which is done by using the appropriate generalizations
of (\ref{l-split}) (see the proof of theorem 
\ref{superenergy-law}, theorem \ref{matter-law} in appendix A and
especially equation (\ref{oth-t})).
The final result is presented next.
\begin{theo}
The following set of equations 
$$
\nb_bT^b_{\ a}=F_a^{\ b}j_b,\ \nb_aj^a=0,
$$
is equivalent to 
\begin{eqnarray}
\pounds_{n}U=-E^aJ_a-2A^aP_a-\frac{4}{3}U\theta
-{\mathcal T}^{ab}\sigma_{ab}-D_aP^a,\ 
\label{poynting-1}\\
\fl\pounds_nP_a=-\varepsilon_{abc}B^bJ^c+E_a\rho+2\varepsilon_{abc}P^b\omega^c
-P_a\theta-A^b(Uh_{ab}+{\mathcal T}_{ab})-D_b{\mathcal T}^b_{\ a},\ 
\label{poynting-2}\\
\pounds_n\rho=-A^aJ_a+\theta\rho+D_aJ^a.
\label{charge-conservation}
\end{eqnarray} 
\label{poynting-theorem}
\end{theo}
Equations (\ref{poynting-1})-(\ref{poynting-2}) are presented 
in basic electrodynamics
books under the heading of the Poynting theorem and they reflect the 
transfer of energy-momentum in a system composed of charged particles and 
electromagnetic fields. 
Indeed, equations (\ref{poynting-1})-(\ref{poynting-2})
provide the well-known physical interpretation of each of the quantities 
appearing in equation (\ref{tem-decomposition}): $U$ is the electromagnetic
energy density, $P^a$ is the Poynting vector and ${\mathcal T}_{ab}$ 
is the stress tensor of the electromagnetic field (see e.g. \cite{JACKSON} 
for detailed explanations about the role of each of these quantities).

We must note at this point that equations (\ref{poynting-1})-(\ref{poynting-2}) 
are usually presented under the assumption that the spacetime is flat and
$n^a$ is chosen
in such a way that all the kinematical quantities vanish. The resulting equations 
can be always obtained {\em locally} in a general spacetime if we recall that we 
can always construct a vector field $n^a$ with the property that all its 
kinematical quantities 
vanish at a prescribed point (equivalence principle). Therefore, we deduce from these
considerations that we can classify the terms which appear in 
(\ref{poynting-1})-(\ref{poynting-2}) 
into two categories: those which contain kinematical quantities 
and those which do not. Terms which do not contain kinematical 
quantities can be regarded 
as representing {\em intrinsic} variations of energy or momentum 
(non-inertial terms) whereas 
terms affected by kinematical quantities can be thought of as depending on 
the observer $n^a$ and
we shall call them {\em inertial terms} in analogy to the inertial forces
introduced in the study of accelerated systems in Newtonian physics. These considerations, 
although elementary, will
play an important role in section \ref{superenergy_balance}.    
    
\subsection{Coupling of the vorticity and the Poynting vector}
If the vector field $n^a$ is hypersurface orthogonal then 
 (\ref{poynting-1})-(\ref{poynting-2}) assume simpler forms 
which can be found in different places in the literature 
\cite{THORNE}. The general form of (\ref{poynting-1}) is written down in 
\cite{MAR-BAS} but to the best of 
our knowledge equation (\ref{poynting-2}) does not seem 
to be present in accessible references. 
Also some of the consequences of (\ref{poynting-2}) do not appear to be widely known. 
To illustrate this fact, consider the inertial 
terms in (\ref{poynting-2}). In ordinary units we find that
the left hand side of (\ref{poynting-2}) is the time variation 
of the momentum density and 
therefore the terms on the right hand side of (\ref{poynting-2}) which 
are coupled to the kinematical quantities can be regarded as inertial forces. 
Indeed equation (\ref{poynting-2}) can be interpreted as an equilibrium 
condition for an electromagnetic system which states that the sum of all 
(inertial and non inertial) forces acting on the system equals zero. 

One of the inertial forces is given by $2\varepsilon_{abc}P^b\omega^c$ or in 
three-vector notation $2\vec{P}\times\vec{\omega}$ with ``$\times$'' 
representing the vector product. If we consider a gyroscope  
then we find that the vorticity $\omega^a$ is related to
the angular velocity of the gyroscope. Therefore we deduce that an inertial force exits on
an uncharged gyroscope when it is placed in a radiative electromagnetic field.  
In SI units this inertial force can be estimated by
\begin{equation}
\vec{F}_g\approx\frac{2V}{c^2}\vec{P}\times\vec{\omega},
\label{precession}
\end{equation}
where $V$ is the volume of the gyroscope.
Thus we conclude that the flux of electromagnetic radiation 
produces an effect on a gyroscope. We must stress at this point that 
this is an observer dependent effect (as it should be because we are dealing 
with an inertial force) which manifests itself in the fact that the angular
velocity of the gyroscope depends on the observer. An effect similar to this 
was pointed out in a  particular  case in \cite{BONNOR} and this 
was latter confirmed in \cite{HERRERA}. 
In the former reference  it was shown that gyroscopes placed in
the spacetime generated by a nonrotating charged magnetic dipole would
precess. As an explanation of this result it was suggested
that the Poynting vector could cause a measurable effect on a gyroscope's
precession and it is conceivable that (\ref{precession}) is related to
this effect in some way.    

\section{Gravitational equations and the Bel tensor}
\label{gravitational-equations}
We start this section by reviewing the well-known formal analogy which exists 
between electromagnetism and gravitation. In this framework the  
Riemann tensor $R_{abcd}$ is taken to be as the gravitational counterpart of the
Faraday tensor $F_{ab}$ and the role of the two Maxwell equations is played by the 
relations
\begin{equation}
\nb_{[a}R_{bc]df}=0,\ \nb_dR_{bpc}^{\ \ \ d}={\mathfrak J}_{pbc},\ 
{\mathfrak J}_{efa}\equiv\nb_{e}R_{af}-\nb_fR_{ae}
\label{bianchi-form}
\end{equation} 
The tensor ${\mathfrak J}_{abc}$ is known as the {\em matter current} and 
can be regarded as the counterpart of the charge current four vector $j^a$. There is an
important difference between electromagnetism and gravitation in that in the 
latter we have an extra set of conditions: the Einstein field equations 
\begin{equation}
G_{ab}={\mathfrak T}_{ab}. 
\label{einstein-equations}
\end{equation}
Here the tensor ${\mathfrak T}_{ab}$ is the energy-momentum tensor of the
system and must be prescribed independently. Clearly any solution of the Einstein 
equations will be a solution of (\ref{bianchi-form}) but the converse need not 
be true. From (\ref{bianchi-form}) we derive the well-known relation (see e.g.
\cite{EDGAR-SENOVILLA})
\begin{eqnarray}
\fl\nb_a\nb^aR_{dcbp}=\nb_b{\mathfrak J}_{dcp}-\nb_p{\mathfrak J}_{dcb}
-2 R_{{dc  }}^{{\  \ ae}}R_{{bape}}^{{    }}
-R_{{b   }}^{{\ a}} R_{{dcpa}}+R_{{dcba}}^{{    }} 
R_{{p   }}^{{\ a}}-2 R_{{cape}}^{{    }}R_{{d\ b }}^{{\ a\ e}}+\nonumber
\\
+2R_{{beca}}^{{    }} R_{{d\ p   }}^{{\ a\ e}},
\label{hyperbolic-equation}
\end{eqnarray}
which can be shown to be a hyperbolic equation
for the Riemann tensor. 
A result due to Lichnerowicz \cite{LICHNEROWICZ} proves
that if the Cauchy data of (\ref{hyperbolic-equation}) 
satisfy (\ref{einstein-equations})
then so does the solution of the hyperbolic equation. Hence, with 
the provision imposed by the Lichnerowicz result, 
we can regard (\ref{bianchi-form}) 
and (\ref{einstein-equations}) as equivalent.

\subsection{Orthogonal splitting of the Riemann tensor}
The orthogonal splitting of the Riemann tensor was first studied in \cite{BEL1}
and since then it has been used in many places.
Define the left, right and double dual of Riemann tensor in the standard 
fashion
$$
\ ^{*}R_{abcd}\equiv\frac{1}{2}\eta_{abpq}R^{pq}_{\ \ cd},\ 
R^{*}_{abcd}\equiv\frac{1}{2}\eta_{pqcd}R_{ab}^{\ \ pq},\ 
\ ^{*}R^*_{abcd}=\frac{1}{2}\eta_{ab}^{\ \ pq}R^*_{pqcd}.
$$
Next we introduce the following spatial tensors \cite{BEL1}
\begin{equation}
Y_{ac}\equiv R_{abcd}n^bn^d,\ Z_{ac}\equiv \ ^{*}R_{abcd}n^bn^d,\ 
X_{ac}\equiv\ ^{*}R^*_{abcd}n^bn^d
\label{riemann-parts}
\end{equation}
The symmetries of Riemann tensor entail the properties
\begin{equation}
X_{(ab)}=X_{ab},\ Y_{(ab)}=Y_{ab},\ Z^a_{\ a}=0.
\label{riemann-parts-prop}
\end{equation}
These tensors contain all the information in the Riemann tensor as is easily checked
by a simple count of their total number of independent components. They also enable
us to find the orthogonal splitting of the Riemann tensor which reads
\begin{eqnarray}
\fl R_{{abcd}}^{{    }}=2n_cn_{[a}Y_{{b]d}}^{{  }}+2h_{{a[d}}^{{  }} X_{{c]b}}^{{ }}
+2n_dn_{[b} Y_{{a]c}}^{{  }}+2n_{[d} Z_{{\ c]}}^{{e }}\varepsilon_{{abe}}^{{   }}
+2n_{[b} Z_{{\ a]}}^{{e }}\varepsilon_{{cde}}+\nonumber\\
+h_{{bd}}^{{  }}\left(h_{{ac}}^{{  }} 
X_{{\  e}}^{{e }}-X_{{ac}}^{{ }}\right)+h_{{bc}}^{{  }}
\left(X_{{ad}}^{{ }}-h_{{ad}}^{{  }} X_{{\   e}}^{{e }}\right).
\label{riemann-split}
\end{eqnarray}
From this expression is easy to get the orthogonal splitting of 
the Ricci tensor which is 
\begin{equation}
R_{ac}=Z^{db} \varepsilon_{cdb}n_a+n_cY_{\ d}^{d } n_a-X_{ac}-Y_{ac}
+n_c Z^{db} \varepsilon_{adb}+h_{ac} X_{\ d}^{d}.
\label{ricci-split}
\end{equation}
The Weyl tensor $C_{abcd}$ has the same algebraic properties as the Riemann 
tensor and in addition it is completely traceless. Therefore 
to find its orthogonal splitting we proceed along the same lines as 
with the Riemann tensor but using different names for 
 the tensors introduced 
in (\ref{riemann-parts}). The precise correspondences are
(in the next equation $X_{ab}$, $Y_{ab}$, $Z_{ab}$ are defined as in
(\ref{riemann-parts}) with the Riemann replaced by the Weyl tensor)  
\begin{equation}
B_{ab}\equiv Z_{ab}=Z_{(ab)},\ E_{ab}\equiv Y_{ab}=-X_{ab},\ E^a_{\ a}=0.
\end{equation}
The tensors $E_{ab}$ and $B_{ab}$ are known as the electric and magnetic 
parts of the Weyl tensor and they completely characterize the former. Equation
(\ref{riemann-split}) becomes for the Weyl tensor
\begin{eqnarray}
\fl C_{{abcd}}^{{    }}=2n_cn_{[a}E_{{b]d}}^{{  }}-2h_{{a[d}}^{{  }}E_{{c]b}}^{{ }}
+2n_dn_{[b}E_{{a]c}}^{{  }}+2n_{[d}B_{{\ c]}}^{{e }}\varepsilon_{{abe}}^{{   }}
+2n_{[b} B_{{\ a]}}^{{e }}\varepsilon_{{cde}}+2h_{{b[d}}^{{  }}E_{{c]a}}^{{ }}.\nonumber\\
\label{riemann-decomposition}
\end{eqnarray}
\subsection{Orthogonal splitting of the matter current}
The orthogonal splitting of ${\mathfrak J}_{abc}$ can be calculated if 
we insert in the last expression of (\ref{bianchi-form}) 
the orthogonal decomposition
of the Ricci tensor (\ref{ricci-split}). 
In this calculation the orthogonal splittings of $\nb_aX_{bc}$, $\nb_aY_{bc}$,
$\nb_aZ_{bc}$, $\nb_a\varepsilon_{bcd}$ must be used (see appendix A for 
the explicit expressions). 
The result is 
\begin{equation}
\mathfrak{J}_{{efa}}^{{   }}=-L_f n_e n_a+L_e n_f n_a+
\tilde{J}_{{fe}}^{{  }} n_a+n_f\overline{J}_{{ea}}^{{    }}-n_e
   \overline{J}_{{fa}}^{{ 
   }}+j_{{efa}}^{{   }},
\label{matter-current-decomposition}
\end{equation}
where 
\begin{eqnarray}
\fl\tilde{J}_{ef}\equiv 2(X_{[f}^{\ a}+Y_{[f}^{\ a})(\sigma_{e]a}+\omega_{e]a})
-2(X^a_{\ a}+Y^a_{\ a})\omega_{ef}+2\varepsilon_{ab[e}D_{f]}Z^{ab},\\
\fl\overline{J}_{{ea}}^{{  }}\equiv 2 Y_{{e }}^{{\ b}} 
\sigma_{{ab}}^{{  }}+X_{{\ b}}^{{b }} \sigma_{{ae}}+h_{{ae}}^{{  }}\left(-\frac{1}{3} X_{{\ b}}^{{b }} 
\theta _{}+X_{{  }}^{{bc}} \sigma _{{bc}}^{{  }}\right)+
(-2 X_{{a }}^{{\ b}}+Y_{{a }}^{{\ b}}) \sigma _{{eb}}^{{  }}+\nonumber\\
\fl+2 Y_{{e }}^{{\ b}} \omega _{{ab}}
-(X_{{\ b}}^{{b }}+2 Y_{{\ b}}^{{b }}) \omega _{{ae}}^{{  }}-
Y_{{a }}^{{\ b}} \omega _{{eb}}^{{  }}+\varepsilon_{{ebc}} (-A^b Z_{{\ a}}^{{c }}
-D_aZ^{{bc}}+D^cZ_{{\ a}}^{{b }})+\nonumber\\
\fl+\varepsilon _{{abc}} (-A^b Z_{{\ e}}^{{c }}+D_eZ_{{  }}^{{bc}})+\frac{1}{3} (X_{{ae}}^{{  }} \theta
_{}-3 (\pounds_nY_{{ae}}^{{  }})),\\
\fl L_e\equiv A_e(X_{{\ a}}^{{a }}+Y_{{\ a}}^{{a }})-A^a (X_{{ea}}^{{  }}+Y_{{ea}}^{{  }})+
2\omega^aZ_{{[ae]}}^{{  }}+
\varepsilon _{{abc}}^{{   }} Z_{{  }}^{{ab}} 
\sigma_{{e}}^{{\ c}}+D_eY_{{\ a}}^{{a }},\\
\fl j_{{efa}}\equiv 2\omega_f Z_{[ea]}+2\omega_e Z_{[af]}-
6\omega^b (h_{{af}}Z_{[eb]}+h_{{ae}}^{{  }}Z_{[bf]})+4\omega_a Z_{[ef]}+\nonumber\\
\fl 2\varepsilon_{{bc[f}}^{{   }} Z_{{  }}^{{bc}} 
\left(\frac{1}{3} h_{{e]a}}^{{  }} \theta _{}+\sigma _{{e]a}}^{{  }}\right)+
2h_{a[f}D_{e]}X_{{\ b}}^{{b }}+2D_{[f}X_{e]a}+2D_{[f}Y_{e]a}^{{  }}.
\end{eqnarray}
From these expressions we deduce 
the properties $\tilde{J}_{[ab]}=\tilde{J}_{ab}$, $j_{[ab]c}=j_{abc}$.

\subsection{The Bel and Bel-Robinson tensors}
Finding a gravitational equivalent of the electromagnetic 
energy-momentum tensor $T_{ab}$ proves to be a delicate issue. 
The reason for this lies in the impossibility of a local 
definition of the {\em gravitational energy-momentum density} 
due to the equivalence principle. Therefore it is clear from the 
very beginning that any tensor qualifying as the gravitational 
counterpart of $T_{ab}$ must represent a physical quantity different
from ``energy-momentum''. If we are unwilling to introduce 
``new quantities'' in physics  then the point of view traditionally adopted
 consists in resorting to quantities defined non locally 
or using {\em pseudo-tensors} (a very good review of the research carried 
out in this direction is \cite{SZABADOS}). However, if we are ready
to deal with a quantity different from ``energy-momentum''  then we
find that it is possible to construct a tensor whose mathematical properties are
similar to the electromagnetic tensor $T_{ab}$ and this is the Bel tensor. 

The Bel tensor was first introduced in \cite{BEL2} in connection with 
the construction of covariant divergences of quantities quadratic in
the Riemann tensor. The original definition given by Bel can be shortened 
to the expression
\begin{equation}
\fl B_{a\ c }^{\ b\ d}\equiv
\frac{1}{2}(R^*_{{apcq}} R^{*\ bpdq}+\ ^{*}R_{{apcq}}\ ^{*}R^{bpdq}+
\ ^{*}R^{*}_{{apcq}}\ ^*R^{*\ bpdq}+R_{apcq} R^{bpdq}),
\label{bel-tensorlong}
\end{equation}
which is 
formally similar to the first equation in (\ref{energy-electromagnetic})
although with more terms due to the fact that the Riemann tensor has 
two blocks of antisymmetric indices.

If we expand the duals in (\ref{bel-tensorlong}) we get
\begin{eqnarray}
\fl B_{abcd}=R_{aecf} R_{b\ d\ }^{\ e\ f}-\frac{1}{2} g_{dc} 
R_{aefp} R_{b  }^{\ efp}-
\frac{1}{2} g_{ba} R_{cefp} R_{d   }^{{\ efp}}+R_{b\ c }^{\ e\ f} R_{dfae}+
\nonumber\\
+\frac{1}{8} g_{{ba}} g_{{dc}}R_{{efph}} R^{{efph}}.
\label{bel-tensor}
\end{eqnarray}
The Bel tensor has a number of remarkable 
mathematical properties which are summarized next.
\begin{theo}
The following statements hold true for the Bel tensor
\begin{enumerate}
\item $B_{abcd}=B_{(ab)(cd)}=B_{cdab}$,\ $B^a_{\ acd}=0$.
\item (Generalized dominant property) If $u_1^a$, $u^a_2$, $u^a_3$, 
$u^a_4$ are arbitrary causal, future directed vectors then 
$B_{abcd}u_1^au_2^bu^c_3u^d_4\geq 0$.
\item $B_{abcd}=0\Longleftrightarrow R_{abcd}=0\Longleftrightarrow
\ \exists$ a timelike vector $u^a$ such that 
$B_{abcd}u^au^bu^cu^d=0$.\label{vanishing-c}
\item \label{point-4} Equation (\ref{bianchi-form}) entails  
\begin{equation}
\fl
\nb_aB^a_{\ bcd}={\mathfrak J}_d^{\ ae}R_{beca}+{\mathfrak J}_c^{\ ae}R_{beda}
-\frac{1}{2}{\rm g}_{cd}{\mathfrak J}^{aef}R_{bfae},\ 
\nb_a{\mathfrak J}_{bc}^{\ \ a}=0.
\label{mattercurrent-conservation}
\end{equation}
\end{enumerate} 
\label{bel-properties}
\end{theo}
The similarity between the mathematical properties of $B_{abcd}$ presented 
in this theorem and those of $T_{ab}$ given by theorem \ref{maxwell-properties} is apparent.
Therefore the Bel tensor fulfills the basic mathematical requirements 
needed for it to be regarded as the gravitational counterpart of the energy-momentum
tensor in electromagnetism.
In vacuum, the Bel tensor acquires a simpler expression which is 
\begin{equation}
T_{abcd}\equiv 
C_{a\ d}^{\ p\ f}C_{bpcf}+C_{a\ c}^{\ p\ f}C_{bpdf}-\frac{1}{8}
{\rm g}_{ab}{\rm g}_{cd}C_{pqrs}C^{pqrs},
\label{bel-robinson}
\end{equation}
where $R_{abcd}=C_{abcd}$ has been used. The tensor $T_{abcd}$ is known 
as the Bel-Robinson tensor \cite{BEL3} and it can be defined in any spacetime,
whether vacuum or not, by means of equation (\ref{bel-robinson}). 
All the properties of theorem \ref{bel-properties} except point 
{\em (\ref{point-4})} are also true for the Bel-Robinson 
tensor with the following changes:
$T_{abcd}$ is totally symmetric and trace-free and 
in point ({\em\ref{vanishing-c}}) the Riemann tensor must be replaced by the Weyl 
tensor. If $R_{ab}=\Lambda {\rm g}_{ab}$ (Einstein space) 
then the covariant divergence of the
Bel-Robinson tensor takes a particularly simple form.
$$
\nb_aT^a_{\ bcd}=0.
$$
A full account of the properties reviewed here of the Bel and  
Bel-Robinson tensors together with their proofs can be found in 
\cite{SUPERENERGY} and \cite{BONILLA}. In the former reference 
a generalization of (\ref{energy-electromagnetic})  
and (\ref{bel-tensorlong}) valid for any tensor is put forward.
Tensors resulting from this generalization are called
{\em superenergy tensors} and they all fulfill the generalized dominant property
({\em generalized dominant superenergy condition}).

What about the physical role of the Bel tensor? This question has been 
addressed many times in the past and no definitive answer exists. 
Bel himself proposed the name of {\em superenergy} for the physical quantity
which might lie behind the Bel tensor (this physical quantity would be
represented by the components of Bel tensor in a suitable frame). 
If we denote by $L$ the basic unit in the geometrized system then 
from the definition of the Bel tensor we deduce
that the physical units of superenergy are 
$L^{-4}$ which can be interpreted
as either energy density squared or energy density per unit area. 
Both interpretations have been researched in the literature and the opinion 
favoring the second interpretation seems to have gained weight.
For a history of the different interpretations of the
Bel tensor which have been studied in the past see \cite{SUPERENERGY} and references
therein.
 
In the case of electromagnetism we have seen that a full understanding of the
physical properties of the electromagnetic energy-momentum tensor can be achieved 
by the Poynting theorem. This theorem is nothing but the orthogonal splitting
of (\ref{e-mconservation}) and the different equations of this splitting 
inform us of the evolution of the different parts of the electromagnetic
energy-momentum tensor. Therefore it is expected that the orthogonal splitting
of equation (\ref{mattercurrent-conservation}) will yield valuable information
about the true physical role of the Bel tensor. The calculation of such an orthogonal 
splitting is accomplished in the forthcoming sections. 

\section{Orthogonal splitting of the Bel-Robinson tensor}
\label{br-splitting}
Before studying the general case of the Bel tensor we 
calculate the orthogonal splitting of the Bel-Robinson tensor.
The different parts of the splitting take simpler forms and they will
give us valuable insights about the general case. To calculate this 
splitting we insert the expression  for the orthogonal splitting of the 
Weyl tensor given by (\ref{riemann-decomposition}) into (\ref{bel-robinson}).
After some computations we get
\begin{equation}
\fl T_{abcd}=Wn_an_bn_cn_d+4{\cal P}_{(a}n_bn_cn_{d)}+6t_{(ab}n_cn_{d)}
+4Q_{(abc}n_{d)}+t_{abcd},
\label{br-decomposition}
\end{equation}
where
\begin{eqnarray}
\fl W\equiv E_{ab}E^{ab}+B_{ab}B^{ab},\  
{\cal P}_a\equiv 2B_{p}^{\ l}E_{ql}\varepsilon_a^{\ pq},\
t_{ab}\equiv W h_{ab}-2(B_a^{\ c}B_{bc}+E_a^{\ c}E_{bc}),\nonumber\\ 
\fl Q_{{cdb}}\equiv h_{{cd}} {\cal P}_b -2 
\left(B_{{da}}^{{  }}E_{{cf}}^{{  }}+B_{{ca}}^{{  }}
E_{{df}}^{{  }}\right) \varepsilon _{b}^{{\ af}},\label{br-parts}\\
\fl t_{{abcd}}^{{    }}\equiv 4(B_{{ab}}^{{  }} 
B_{{cd}}^{{  }}+E_{{ab}}^{{  }} E_{{cd}}^{{  }})
-h_{{cd}} t_{{ab}}^{{  }}+2h_{{b(d}} 
t_{{c)a}}+2h_{{a(d}}t_{{c)b}}-h_{{ab}}^{{  }} t_{{cd}}+\nonumber\\
+W(h_{ab}h_{cd}-2h_{a(c} h_{d)b})\nonumber.
\end{eqnarray}
Some of these quantities have been obtained before and
have found diverse applications. The scalar $W$ (superenergy
density) and the spatial vector ${\cal P}^a$, called the super-Poynting 
vector, were first used in \cite{BEL4} to define intrinsic {\em radiation 
states} in gravitation theory (see section \ref{superenergy_balance} for 
more details about this) and the tensor $t_{ab}$ was used in 
\cite{BONILLA2} to show the causal propagation of gravity in
vacuum (also the role of $t_{ab}$ in the definition of radiation states
was discussed in this reference).  
We establish next the basic algebraic properties of these quantities
\begin{prop}
The following basic algebraic properties hold 
\begin{enumerate}
\item $t_{(ab)}=t_{ab}$, $Q_{(abc)}=Q_{abc}$, $t_{(abcd)}=t_{abcd}$,
\label{pointfirst}
\item $t^a_{\ a}=W\geq 0$, $Q^a_{\ ab}={\cal P}_b$, $t^a_{\ abc}=t_{bc}$,
\label{pointsecond}
\item $Q_{abc}$ and $t_{abcd}$ contain all the information about 
the Bel-Robinson tensor.
\label{pointthird}
\end{enumerate}
\label{algebraic-properties}
\end{prop}
\P Points ({\em \ref{pointfirst}}) and ({\em \ref{pointsecond}}) 
can be proven directly from 
the tensor expressions given in (\ref{br-parts}) but it is far 
more easier to use (\ref{br-decomposition}) and write each part of the 
decomposition in terms of the Bel-Robinson tensor $T_{abcd}$. The result is
\begin{eqnarray}
\fl W=T_{abcd}n^an^bn^cn^d,\ 
{\cal P}_e=-T_{ebcd}h_{\ a}^{e}n^bn^cn^d,\ 
t_{ab}=T_{pqrs}h^p_{\ a}h^q_{\ b}n^rn^s,\label{br-inverse}\\ 
\fl Q_{abc}=-T_{pqrs}h^p_{\ a}h^q_{\ b}h^r_{\ c}n^s,\ 
t_{abcd}=T_{pqrs}h^p_{\ a}h^q_{\ b}h^r_{\ c}h^s_{\ d}.
\label{br-inverse1}
\end{eqnarray} 
The symmetries expressed in point ({\em \ref{pointfirst}}) 
are now a consequence of the total symmetry of $T_{abcd}$.
Point ({\em \ref{pointsecond}}) is straightforward either from (\ref{br-parts})
or from (\ref{br-inverse})-(\ref{br-inverse1}) and the complete tracelessness
of the Bel-Robinson tensor. 
Thus given $t_{abcd}$ and $Q_{abc}$ it is evident from their algebraic 
properties that we recover the remaining parts of the orthogonal 
decomposition of the Bel-Robinson tensor 
which proves point ({\em \ref{pointthird}}).
\N  
\begin{Remark}\em
We can obtain an independent proof of point ({\em \ref{pointthird}}) of 
the previous proposition if we count the number of independent components of 
$Q_{abc}$ and $t_{abcd}$ and compare their sum with the number of total
independent components of $T_{abcd}$. 
The respective numbers are 
\begin{eqnarray*}
\mbox{number of independent components of $T_{abcd}$}=25,\\
\mbox{number of independent components of $t_{abcd}$}=15,\\
\mbox{number of independent components of $Q_{abc}$}=10,\\
10+15=25.
\end{eqnarray*}
\end{Remark}
\begin{prop}
\item $t_{abcd}=0$ $\Longleftrightarrow$ $t_{ab}=0$ 
$\Longleftrightarrow$ $W=0$ $\Longleftrightarrow$ $C_{abcd}=0$,\ 
\label{weyl-zero}
\end{prop}
\P
From proposition 
\ref{algebraic-properties} we deduce 
$t_{abcd}=0$ $\Longrightarrow$ $t_{ab}=0$
$\Longrightarrow$ $W=0$. To prove the converse, let us assume that  
$W=0$. In this case the first equation of (\ref{br-inverse}) 
implies 
$$
T_{abcd}n^an^bn^cn^d=0.
$$
Combining this with point ({\em \ref{vanishing-c}}) of theorem 
\ref{bel-properties} applied to the Bel-Robinson tensor we deduce 
$C_{abcd}=0$. Trivially, $C_{abcd}=0$ implies $T_{abcd}=0$ 
and thus $t_{ab}$, $t_{abcd}$ vanish as well.\N

The importance of this result lies in the fact that evaluation of any of 
the quantities $W$, $t_{ab}$, $t_{abcd}$ enables a observer
represented by the unit timelike vector $n^a$ to decide if the 
purely gravitational part of the Riemann tensor (or the Riemann
tensor itself if we are in a vacuum spacetime) is present or not.
Also the variation of these quantities along the integral curves
of $n^a$ should give a measure of 
how the Weyl tensor changes for this observer. We will turn back 
to this important point in section \ref{superenergy-evolution}.   

\subsection{Canonical forms for the different Petrov types}
\label{petrov-forms}
We can obtain more interesting properties of 
the quantities introduced in (\ref{br-parts}) if we
set up a suitable orthonormal frame.
Such a frame arises in the calculation of the  
canonical forms which $E_{ab}$ and $B_{ab}$ take for the different
{\em Petrov types}. These canonical forms are reviewed in appendix B and 
we refer the reader to this appendix for more details. 
The results presented in this subsection are algebraic in nature 
and should be understood as formulated in the tangent space of a point.
\begin{prop}
The tensor $Q_{abc}$ vanishes if and only if $E_{ab}$, $B_{ab}$
are linearly dependent.
\label{poynting-like}
\end{prop}
\P From (\ref{br-parts}) it is easy to show that $Q_{abc}$ is 
zero if $E_{ab}$ and $B_{ab}$ are linearly dependent. 
Now, if $Q_{abc}=0$ then from point ({\em \ref{pointsecond}}) of proposition
\ref{algebraic-properties} we get ${\cal P}^a=0$. This last condition
can be re-written in the form
\begin{equation}
E_a^{\ r}B_{r b}-E_b^{\ r}B_{r a}=0,
\label{e-b}
\end{equation}
from which we conclude that the endomorphisms represented by $E^a_{\ b}$, 
$B^a_{\ b}$ commute. This is only possible for Petrov types I and D
as can be easily checked using the canonical forms of appendix B 
(alternatively, two symmetric endomorphisms have a common 
basis of eigenvectors if and only if they commute).
For Petrov type D trivially $E_{ab}$ and $B_{ab}$ are linearly dependent, 
so we will assume that the spacetime is of Petrov type
I. In the orthonormal frame of (\ref{type-I}) we find that the only 
nonvanishing component of $Q_{abc}$ is  
$$
Q_{123}=-2(B_{11}E_{22}-B_{22}E_{11}),
$$
and hence $Q_{123}=0$ implies $B_{11}E_{22}=B_{22}E_{11}$ 
from which we deduce from (\ref{type-I}) that $E_{ab}$ and 
$B_{ab}$ are linearly dependent (recall that 
$E_{11}+E_{22}+E_{33}=B_{11}+B_{22}+B_{33}=0$).\N 

From this result we deduce that $Q_{abc}$ resembles in its 
mathematical properties the electromagnetic Poynting vector.
We will see later that if we are to study the radiation of superenergy
then $Q_{abc}$ (or any equivalent tensor thereof) will take over the
role of the Poynting vector.

Proposition \ref{poynting-like} admits the following corollary.
\begin{coro}
\ 
\begin{enumerate} 
\item $Q_{abc}\neq 0$ $\Longrightarrow$ Petrov type is either II, III, N or I. 
\item If Petrov type is II, III, or N $\Longrightarrow$ $Q_{abc}\neq 0$.
\item Petrov type D is the only type in which $Q_{abc}$ always vanishes. 
\end{enumerate}
\end{coro}

\begin{prop}
\ The following algebraic properties hold 
\begin{enumerate}
\item \label{point1-canonical} For Petrov type III we have
\begin{eqnarray*}
\fl t_{ab}=\frac{1}{2}h_{ab}W-\frac{2{\cal P}_a{\cal P}_b}{W},\ 
Q_{abc}=3 h_{(bc}{\cal P}_{a)}-\frac{16{\cal P}_a{\cal P}_b{\cal P}_c}{W^2},\\ 
\fl t_{abcd}=-\frac{64{\cal P}_a{\cal P}_b{\cal P}_c{\cal P}_d}{W^3}
+\frac{12}{W}h_{(ab}{\cal P}_{c}{\cal P}_{d)},\ 
{\cal P}_a\neq 0,\ {\cal P}_a{\cal P}^a=\frac{W^2}{4}.
\end{eqnarray*}
The two independent principal null directions of the Weyl tensor 
(see e.g. \cite{MCCALLUM}) 
can be calculated explicitly yielding
\begin{equation}
k_1^a=-{\cal P}^a+\frac{1}{2}n^aW,\ 
k_2^a={\cal P}^a+\frac{1}{2}n^aW.
\end{equation}
\item\label{point2-canonical} For Petrov type N we have 
\begin{eqnarray*}
\fl t_{ab}=\frac{{\cal P}_a{\cal P}_b}{W},\ 
Q_{abc}=\frac{{\cal P}_a{\cal P}_b{\cal P}_c}{W^2},\ 
t_{abcd}=\frac{{\cal P}_a{\cal P}_b{\cal P}_c{\cal P}_d}{W^3},\ 
{\cal P}_a\neq 0,\ {\cal P}_a{\cal P}^a=W^2. 
\end{eqnarray*}
In this case the only independent principal null direction of the Weyl tensor is
\begin{equation}
k^a\equiv Wn^a+{\cal P}^a.
\end{equation}
\end{enumerate}
\label{t-type}
\end{prop}
\P The proof of this result consists in 
using the canonical forms for Petrov types 
III and N written in appendix B to find canonical forms for
$t_{ab}$, ${\cal P}^a$, $Q_{abc}$ and $t_{abcd}$. 
These canonical forms lead then 
to the expressions presented in points {\em (\ref{point1-canonical})} and 
{\em (\ref{point2-canonical})}. We detail next this procedure for each of 
the Petrov types.  

\noindent
{\em -- Petrov type III:} using the frame of (\ref{type-III}) we get 
$$
-Q_{133}=-Q_{122}=Q_{111}=2(B_{12}^2+E_{12}^2),\ 
{\cal P}_1=-2(E_{12}^2+B_{12}^2),\ 
t_{22}=t_{33}=2(E_{12}^2+B_{12}^2).
$$
with all the other components of $Q_{abc}$, ${\cal P}^a$, $t_{ab}$ being zero.
From these expressions we deduce
\begin{equation}
\fl Q_{abc}=h_{bc}{\cal P}_a+h_{ac}{\cal P}_b+h_{ab}{\cal P}_c
-\frac{{\cal P}_a{\cal P}_b{\cal P}_c}{B_{12}^2+E_{12}^2},\ 
t_{ab}=2(E_{12}^2+B_{12}^2)h_{ab}-\frac{{\cal P}_a{\cal P}_b}
{2(E_{12}^2+B_{12}^2)},
\label{stage-1}
\end{equation}
and using point {\em (\ref{pointsecond})} of proposition
\ref{algebraic-properties}  
we conclude 
$$
B_{12}^2+E_{12}^2=\frac{\sqrt{{\cal P}_a{\cal P}^a}}{2},\ 
{\cal P}_a{\cal P}^a=\frac{W^2}{4}.
$$
Replacing this back in (\ref{stage-1}) we obtain the expressions sought
for $t_{ab}$ and $Q_{abc}$. Inserting the values just found for $t_{ab}$
in the formula for $t_{abcd}$ of (\ref{br-parts}) yields
$$
t_{abcd}=4(B_{ab}B_{cd}+E_{ab}E_{cd})
+\frac{2h_{ab}{\cal P}_c{\cal P}_d}{W}-
\frac{4{\cal P}_b}{W}h_{a(d}{\cal P}_{c)}+
\frac{2{\cal P}_a}{W}(h_{cd}{\cal P}_b-h_{bd}{\cal P}_c
-h_{bc}{\cal P}_d). 
$$ 
Again using the canonical forms of (\ref{type-III}) we transform the 
term $4(B_{ab}B_{cd}+E_{ab}E_{cd})$ into 
$$
-\frac{64{\cal P}_a{\cal P}_b{\cal P}_c{\cal P}_d}{W^3}+
\frac{4}{W}({\cal P}_c(h_{bd}{\cal P}_a+h_{ad}{\cal P}_b)+
(h_{bc}{\cal P}_a+h_{ac}{\cal P}_b){\cal P}_d).
$$ 
Combining the last two equations we find the expression for 
$t_{abcd}$ given in the proposition. It is now a simple calculation 
to check that the vectors $k_1^a$ and $k_2^a$ are indeed null and 
that they fulfill the properties
$$
T_{abcd}k_1^ak_1^bk_1^ck_1^d=0,\
T_{abcd}k_2^ak_2^bk_2^ck_2^d=0 
$$
which implies that $k_1^a$ and $k_2^a$ are the Weyl tensor principal 
null directions (see \cite{PENROSE-RINDLER} p. 328).

\noindent 
{\em -- Petrov type N:} In this case, we obtain in the frame of
(\ref{type-N}) 
$$
Q_{111}={\cal P}_1=-t_{11}=-4(B_{22}^2+E_{22}^2),
$$
with the other components vanishing. Hence
$$
Q_{abc}=\frac{{\cal P}_a{\cal P}_b{\cal P}_c}{16(B_{22}^2+E_{22}^2)^2},\ 
t_{cd}=\frac{{\cal P}_c{\cal P}_d}{4(B_{22}^2+E_{22}^2)}\Rightarrow
B_{22}^2+E_{22}^2=\frac{\sqrt{{\cal P}_a{\cal P}^a}}{4}=\frac{W}{4}.
$$
Similarly, working in the canonical frame we obtain that the only nonvanishing
component of $t_{abcd}$ is
$$
t_{1111}=4(B_{22}^2+E_{22}^2).
$$
Combining the previous pair of equations the expressions of point 
{\em (\ref{point2-canonical})} follow. Also it is a simple matter to check that
$k^a$ is null and that $T_{abcd}k^ak^bk^ck^d=0$.\N

An important result of this proposition is that for Petrov types 
III and N the Bel-Robinson tensor is characterized by just two independent 
quantities which are $W$ and ${\cal P}^a$ and thus 
we can say that the number of algebraically independent
components of the Bel-Robinson tensor is two for these Petrov types. 
This is not true of the other Petrov types and therefore some conclusions 
drawn from considerations involving type III and N might not carry over to 
other Petrov types. An example of this is the definition and study of 
gravitational radiation using the Bel-Robinson tensor where, traditionally,
a nonvanishing vector ${\cal P}^a$ for any observer $n^a$ has been regarded
as an {\em intrinsic state} of gravitational radiation \cite{BEL4}    
(see definition \ref{bel-definition}).
We will see in section \ref{superenergy-evolution} that 
this condition is not general enough and indeed in certain Petrov 
type I spacetimes
we can still speak of an intrinsic state of gravitational 
radiation with ${\cal P}^a$
being zero.

\section{Orthogonal splitting of the Bel tensor}
\label{bel-splitting}
The orthogonal splitting of the Bel tensor is obtained by replacing the expression 
for the Riemann tensor given by (\ref{riemann-split}) in (\ref{bel-tensor}) with 
the result
\begin{eqnarray}
\fl B_{abcd}=\overline{W}n_an_bn_cn_d+4\overline{{\cal P}}_{(a}n_bn_cn_{d)}+
2n_{(a}\overline{Q}_{b)cd}+2n_{(d}\overline{Q}_{c)ab}+\nonumber\\
\fl+\overline{t}_{ab}n_cn_d+\overline{t}_{cd}n_an_b
+4n_{(a}t^*_{b)(c}n_{d)} 
+\overline{t}_{abcd}.
\label{beltensor-split}
\end{eqnarray}
Each of the spatial parts of the Bel tensor is defined as follows
\begin{eqnarray}
\fl \overline{W}\equiv\frac{1}{2}(X_{ab}X^{ab}+Y_{ab}Y^{ab})+Z_{ab}Z^{ab},\ 
\overline{\cal P}_a
\equiv\varepsilon_{abc}(Y_d^{\ c}Z^{bd}-X_{d}^{\ c}Z^{db}),\nonumber\\
\fl \overline{t}_{cd}
\equiv h_{cd}\overline{W}-X_c^{\ a}X_{da}-Y_c^{\ a}Y_{da}-Z_{ac}Z^{a}_{\ d}
-Z_c^{\ a}Z_{da},\nonumber\\
\fl t^*_{bd}\equiv 2X_{(d}^{\ \ a}Y_{b)a}-X_{bd}Y^a_{\ a}-Y_{bd}X^a_{\ a}
+h_{bd}(-X_{{  }}^{{ac}}
   Y_{{ac}}^{{  }}+Z_{{ 
   }}^{{ac}} Z_{{ac}}^{{ 
   }}+X_{{\ a}}^{{a }} Y_{{\ \ c}}^{{c }})
-\nonumber\\
-Z_{{\ b}}^{{a }}Z_{{ad}}^{{  }}-Z_{{b   }}^{{\ a}} Z_{{da}}^{{  }}\nonumber\\
\fl\overline{Q}_{{bcd}}\equiv h_{{cd}}^{{  }}
   \overline{\cal P}_b+2Z_{{\  (d}}^{{a }} 
\left(-Y_{{c)}}^{{\ e}} \varepsilon_{{bae}}^{{   }}
+\varepsilon_{{c)ba}}^{{   }}X_{{\  e}}^{{e }}+
\varepsilon_{{c)ae}}X_{{b}}^{{\ e}}\right)+\nonumber\\
\fl +2Z_{{  }}^{{ae}}\left(h_{{b(c}}(\varepsilon _{{d)ae}}X_{{\ f}}^{{f }}
-\varepsilon_{{d)af}}^{{   }}X_{{e }}^{{\ f}})-X_{{e(d}}
\varepsilon_{{c)ba}}+h_{{cd}} 
X_{{a}}^{{\ f}} \varepsilon_{{bef}}-X_{{b(d}}
\varepsilon_{{c)ae}}\right)\nonumber.   
\end{eqnarray}
The expression for $\overline{t}_{abcd}$ is a bit long and is omitted (its
explicit form is not needed in this paper).
\begin{prop}
The tensors $\overline{t}_{ab}$, 
$t^*_{ab}$, $\overline{Q}_{abc}$ and $\overline{t}_{abcd}$ 
satisfy the following basic algebraic properties
\begin{eqnarray*}
\fl\overline{t}_{(ab)}=\overline{t}_{ab},\ 
t^*_{(ab)}=t^*_{ab},\ \overline{Q}_{a(bc)}=\overline{Q}_{abc},\ 
\overline{t}_{(ab)cd}=\overline{t}_{abcd}=\overline{t}_{cdab},\ 
\overline{t}^a_{\ a}=\overline{W},\ 
\overline{Q}_{b\ a}^{\ a}
=\overline{{\cal P}}_b,\\
\fl\overline{t}^a_{\ abc}=\overline{t}_{bc}
\end{eqnarray*}
\label{bel-parts-properties}
\end{prop} 
\P These properties can be proven from a direct computation using 
 the definitions of $\overline{t}_{ab}$, $t^*_{ab}$, 
$\overline{Q}_{abc}$ and $t_{abcd}$ given above but this results
in involved calculations even when done by computer. A simpler 
procedure is to start with (\ref{beltensor-split}) and  
derive the relations  
\begin{eqnarray}
\overline{W}=B_{abcd}n^an^bn^cn^d,\ 
\overline{\cal P}_a=-B_{pbcd}h^p_{\ a}n^bn^cn^d,\\ 
\fl\overline{t}_{ab}=B_{pqcd}n^cn^dh^p_{\ a}h^q_{\ b},\ 
t^*_{ab}=B_{rpsq}n^pn^qh^r_{\ a}h^s_{\ b},\ 
\overline{Q}_{abc}=-B_{pqrs}n^ph^q_{\ a}h^r_{\ b}h^s_{\ c},\\
\overline{t}_{abcd}=B_{pqrs}h^p_{\ a}h^q_{\ b}h^r_{\ c}h^s_{\ d}.
\end{eqnarray}
From these relations and the properties of the Bel tensor it is 
straightforward to prove the proposition. \N
\begin{Remark}\em
An important consequence of the algebraic properties 
presented in this last result is that 
$\overline{t}_{abcd}$, $\overline{Q}_{abc}$ and $t^*_{ab}$ 
contain all the information about the Bel tensor. As 
we did in the case of the Bel-Robinson tensor we can count the number of 
independent components of these tensors and check that they add up to 
the number of independent components of the Bel tensor
\begin{eqnarray*}
\mbox{number of independent components of $B_{abcd}$}=45,\\
\mbox{number of independent components of $\overline{t}_{abcd}$}=21,\\
\mbox{number of independent components of $\overline{Q}_{abc}$}=18,\\
\mbox{number of independent components of $t^*_{ab}$}=6.\\
21+18+6=45.
\end{eqnarray*}
\end{Remark}
\begin{prop} 
$$
\overline{W}=0\Longleftrightarrow 
\overline{t}_{ab}=0\Longleftrightarrow
\overline{t}_{abcd}=0\Longleftrightarrow
R_{abcd}=0,\ (\mbox{no superenergy $\Longleftrightarrow$ no gravitation})
$$
\label{riemann-flat}
\end{prop}
\P If $R_{abcd}$ vanishes then so does $B_{abcd}$ and trivially  
$\overline{W}=0$, $\overline{t}_{ab}=0$, $\overline{t}_{abcd}=0$.
Assume now that $\overline{W}$ is zero. In that case point 
({\em \ref{vanishing-c}}) of theorem \ref{bel-properties} entails
$R_{abcd}=0$ thus proving the desired result.\N 

We finish this section by pointing out that whenever the Bel and Bel-Robinson 
tensors are equal then
we deduce the relations 
$$
t^*_{ab}=\overline{t}_{ab}=t_{ab},\ \overline{Q}_{abc}=Q_{abc},\ 
\overline{t}_{abcd}=t_{abcd},
$$
from which we conclude that $\overline{W}=W$, 
$\overline{\cal P}_a={\cal P}_a$. The Bel and the Bel-Robinson tensors are
equal if and only if $R_{ab}=0$ (see Corollary 6.1 of \cite{SUPERENERGY}).

\section{Dynamical laws of superenergy}
\label{superenergy-evolution}
In this section we present the most important result of this paper 
which is the orthogonal splitting of (\ref{mattercurrent-conservation}).
As explained before this result is analogous to 
(\ref{poynting-1})-(\ref{charge-conservation}) and this analogy will enable
us to extract some interesting conclusions as to the interpretation of
certain parts of the orthogonal splitting of the Bel tensor.

Before presenting the results we should make some remarks concerning the 
calculations. In order to work out the orthogonal splitting of 
(\ref{mattercurrent-conservation}) 
neither (\ref{bianchi-form}), nor its orthogonal splitting 
is needed. This is similar 
to electromagnetism, where the Maxwell equations are not needed to
obtain (\ref{poynting-1})-(\ref{charge-conservation}). The orthogonal 
splitting of (\ref{mattercurrent-conservation}) is calculated by
inserting the orthogonal splitting of each of the quantities appearing
in this equation (the Bel tensor, the Riemann tensor and the matter current) 
and then using the orthogonal splitting of the different terms which appear
in the resulting expressions. Here we only 
provide the final expressions referring the reader to appendix A for
more details about the intermediate steps in the calculations.

\begin{theo}[Dynamical laws of superenergy]
The equation 
$$\nb_aB^a_{\ bcd}={\mathfrak J}_d^{\ ae}R_{beca}
+{\mathfrak J}_c^{\ ae}R_{beda}
-\frac{1}{2}{\rm g}_{cd}{\mathfrak J}^{aef}R_{bfae}
$$
is equivalent to the following set of expressions
\begin{eqnarray}
\fl\pounds_n\overline{\cal P}_c+D_at^{*a}_{\ \ c}+
\left(\overline{\cal P}_c+\frac{2
   \overline{Q}_{{\ ca}}^{{a}}}{3}\right) 
\theta_{}^{}-j_{{c}}^{{\ ae}} Y_{{ae}}^{{  }}+2
   \sigma _{{ae}}^{{  }}
   \overline{Q}_{{\ c }}^{{a\ e}}-Z_{{\ e}}^{{f }}\varepsilon_{{caf}}^{{   }}
   \overline{J}_{{  }}^{{ae}}+\nonumber\\
+2\omega_{{  }}^{{eb}}
   \left(\overline{Q}_{{ecb}}-h_{{ce}}\overline{\cal P}_b\right)
+A^a\left(h_{{ca}}^{{  }}
   \overline{W}_{}^{}+\overline{t}_{{ca}}^{{  }}+2
   t^*_{{ca}}\right)=0,\label{poynting-flow}
\end{eqnarray}
\begin{eqnarray}
\fl D_at^{*a}_{\ \ c}-D_a\overline{t}^a_{\ c}=
L^a Y_{{ac}}^{{  }}+
2Z_{{\ [a}}^{{e }} \varepsilon_{{b]ce}}
\overline{J}_{{ }}^{{ab}}-\frac{1}{2}Z_{{\ c}}^{{a }} 
\varepsilon_{{abe}}^{{   }} \tilde{J}_{{ }}^{{be}}+\nonumber\\
+j_{{   }}^{{abe}}
   \left(h_{{ac}}^{{  }}
   (-X_{{\ d}}^{{d }}
   h_{{be}}^{{  }}+X_{{be}}^{{}}+Y_{{be}}^{{ 
   }})-h_{{ae}}^{{  }}
   X_{{bc}}^{{  }}\right),\label{tstar-constraint}
\end{eqnarray}
\begin{eqnarray}
\fl\pounds_n\overline{t}_{cd}+D_a\overline{Q}^a_{\ cd}+
\omega^{af}\Omega_{afcd}+\sigma^{af}\Sigma_{afcd}+
\frac{2}{3}(\overline{t}_{cd}+t^*_{cd})\theta+
2A^a(h_{a(d}\overline{\cal P}_{c)}+\overline{Q}_{acd})-\nonumber\\
\fl-j_{{   }}^{{fhb}}
\left(\frac{1}{2} \varepsilon_{{afh}}^{{   }} h_{{cd}}
+2h_{{f(d}}\varepsilon_{{c)ah}}\right)
   Z_{{\ b}}^{{a  }}+\left(h_{{cd}}Y_{{ae}}-h_{{da}}Y_{{ce}}-h_{{ca}} 
Y_{{de}}^{{  }}\right)
   \overline{J}^{{ae}}=0,\label{tb-flow}\\
\fl\pounds_nt^*_{bd}+
D_a\overline{Q}^a_{\ bd}+
\omega^{af}\Omega^*_{afbd}+\sigma^{af}\Sigma^*_{afbd}+
\frac{1}{3}
   \left(\overline{t}_{{bd}}+2
   t^*_{{bd}}+\overline{t}_{{b\ da}}^{{\ a  }}\right) \theta+\nonumber\\
\fl A^a \left(2h_{{a(d}}^{{  }}
   \overline{\cal P}_{b)}+\overline{Q}_{abd}+\overline{Q}_{(bd)a}\right)
-j^{cfh}\varepsilon_{ha(d}h_{b)f}Z^a_{\ c}
-L^a \varepsilon_{ae(d}Z_{\ b)}^{e}+
\tilde{J}^{ae}h_{a(d}Y_{b)e}+\nonumber\\
\fl+\left(h_{a(d}X_{b)e}+h_{e(d}(X_{b)a}-h_{b)a}X^p_{\ p})-
h_{bd}(X_{ae}-h_{ae}X^p_{\ p})-h_{ae}X_{bd} 
\right)
   \overline{J}_{{  }}^{{ae}}=0,\label{ts-flow}
\end{eqnarray}
\begin{eqnarray}
\fl\pounds_{n}\overline{Q}_{bcd}+D_a\overline{t}^a_{\ bcd}+
\omega^{af}\Pi_{afbcd}+\sigma^{af}\Delta_{afbcd}+
\theta\overline{Q}_{(bcd)}+A^a(\overline{t}_{abcd}+h_{ab}\overline{t}_{cd}+
2h_{a(d}t^*_{c)b})+\nonumber\\
\fl+\tilde{J}^{ef}Z^a_{\ b}\left(\frac{1}{2} 
\varepsilon_{{aef}}^{{   }}h_{{cd}}+2h_{{e(d}}^{{  }} 
\varepsilon_{{c)af}}^{{  }}\right)
-\overline{J}^{ae}\varepsilon_{bef}(h_{cd}Z^f_{\ a}-2h_{a(d}Z^f_{\ c)})+
\nonumber\\
+L^a(h_{cd}Y_{ab}-2h_{a(d}Y_{c)b})
+j^{aef}H_{aefbcd}=0,\label{qb-flow}
\end{eqnarray}
where
\begin{eqnarray}
\fl\Sigma_{afcd}\equiv 
-2h_{{a(d}}^{{  }}\overline{t}_{{c)f}}+2h_{{a(d}}t^*_{{c)f}}
+\overline{t}_{{cdaf}},\
\Omega_{afcd}\equiv 
-2h_{{a(d}}^{{  }}
   \overline{t}_{{c)f}}-2h_{{a(d}} t^*_{{c)f}},\label{Sigma}\\
\fl\Sigma^*_{afbd}\equiv\overline{t}_{abdf}+
h_{a(d}(\overline{t}_{b)f}-t^*_{b)f}),\ 
\Omega^*_{afbd}\equiv -h_{a(d}(\overline{t}_{b)f}+3t^*_{b)f}),\nonumber\\
\fl\Pi_{afbcd}\equiv-2 \overline{Q}_{{fcd}}^{{  
   }} h_{{ab}}^{{ 
   }}- 4\overline{Q}_{{(bf)(c}}^{{  
   }}h_{{d)a}}^{{  }},\ 
\Delta_{afbcd}\equiv
4\overline{Q}_{{[fb](c}}h_{{d)a}}^{{  }},\nonumber\\
\fl H_{aefbcd}\equiv 2h_{ab}h_{e(d}X_{c)f}+2h_{ef}h_{a(d}X_{c)b}
+2h_{ad}h_{c[b}X_{f]e}+4h_{c[d}h_{a][f}X_{b]e}+\nonumber\\
+X^h_{\ h}(h_{ab}h_{cd}h_{ef}-2h_{ab}h_{c(f}h_{e)d}-2h_{ef}h_{a(d}h_{c)b})
\nonumber
\end{eqnarray}
\label{superenergy-law}
\end{theo}
\P See appendix A.\N
\begin{theo}[Matter current conservation]
The equation $\nb_a{\mathfrak J}_{bc}^{\ \ a}=0$ is equivalent to
the expressions
\begin{eqnarray}
\fl\pounds_nL_p=A^q(\tilde{J}_{pq}-\overline{J}_{pq})
+\frac{\theta}{3}(j_{p\ q}^{\ q}-2L_p)+(L_bh_{pa}-j_{apb})\sigma^{ab}+
(j_{apb}+L_bh_{pa})\omega^{ab}-\nonumber\\
-D_q\overline{J}_p^{\ q},\label{lb-flow}\\
\fl\pounds_n\tilde{J}_{bp}=D_qj_{bp}^{\ \ q}+
\frac{1}{3}
   \left(2\overline{J}_{[bp]}-\tilde{J}_{bp}\right)
   \theta_{}+A^q\left(2h_{q[p}L_{b]}-j_{bpq}\right)+\nonumber\\
+2\sigma^{ac}\left(h_{{a[p}}^{{  }}\overline{J}_{b]c}
+h_{a[p}\tilde{J}_{{b]c}}\right)+2\omega_{{  }}^{{ac}}
   \left(-h_{{a[p}}^{{  }}\overline{J}_{b]c}+h_{a[p}\tilde{J}_{{b]c}}\right).
\label{jtilde-flow}
\end{eqnarray}
\label{matter-law}
\end{theo}
\P Again see appendix A.\N
\begin{Remark}\em
Equations (\ref{poynting-flow})-(\ref{qb-flow}) and 
(\ref{lb-flow})-(\ref{jtilde-flow}) can be regarded as the gravitational
counterpart of (\ref{poynting-1})-(\ref{charge-conservation}). They
form an {\em inhomogeneous} evolution system for the variables 
$\overline{\cal P}_a$,
$\overline{t}_{ab}$, $t^*_{ab}$, $\overline{Q}_{abc}$, $L_a$ and 
$\tilde{J}_{ab}$. 
The inhomogeneous part (source) of each equation consists of 
those terms which contain neither kinematical quantities
nor spatial covariant derivatives. These terms play the 
same role as $-E_aJ^a$ in (\ref{poynting-1}) (power lost by the 
charge flux) and $\varepsilon_{acb}B^bJ^c+E_a\rho$
(change of momentum due to charges) in (\ref{poynting-2}).  
We also find that no 
expressions for $\pounds_nt_{abcd}$, $\pounds_n\overline{J}_{ab}$,
$\pounds_nj_{abc}$  are supplied by the orthogonal splitting 
of (\ref{mattercurrent-conservation}) and in fact only by 
using the full content of (\ref{bianchi-form}) can such 
expressions be found.
\end{Remark}

\begin{Remark}\em
The evolution equations of theorems \ref{superenergy-law} and 
\ref{matter-law} are written in such a way that the coupling of the 
kinematical quantities to the different parts of the orthogonal decomposition
of the Bel tensor and the matter current is manifest. Note also that in these 
equations we can find terms which do not contain kinematical quantities. 
As the kinematical quantities can be always set to zero at a a given point
by choosing a suitable vector field $n^a$ we deduce that 
any term containing explicitly a kinematic quantity is {\em observer
dependent} and it will play a similar role as the inertial terms in
equations (\ref{poynting-1})-(\ref{charge-conservation}) found for 
electromagnetism. 
\label{coupling}
\end{Remark}
Taking the trace of (\ref{tb-flow}) we find 
\begin{equation}
\fl\pounds_n\overline{W}+D_a\overline{\cal P}^a+
\sigma^{ae}(\overline{t}_{ae}+2t^*_{ae})
+\frac{2\theta}{3}(t^{*a}_{\ \ a}+2\overline{W})+
\frac{1}{2}\varepsilon_{afb}j^{fb}_{\ \ e}Z^{ae}+Y^{ae}\overline{J}_{ae}+
4A^a\overline{\cal P}_a=0.
\label{w-flow}
\end{equation}
Equations similar to this one have been used in different places of 
the literature principally with the 
aim of controlling the evolution of the scalar $\overline{W}$
\cite{BRUHAT1,KLAINERMAN}. 

\subsection{Dynamical laws of superenergy in vacuum}
Theorem \ref{superenergy-law} assumes a far more simpler 
form in vacuum because the covariant divergence 
of the Bel tensor takes the simpler form $\nb_aT^a_{\ bcd}=0$. 
The specific result in this case is given in the next theorem.
\begin{theo}
The equation 
$$
\nb_aT^a_{\ bcd}=0,
$$
is equivalent to the following set of expressions
\begin{eqnarray}
\fl\pounds_nt_{cd}=-2A^a (h_{{a(d}}{\cal P}_{c)}+ Q_{{cda}})+
4\omega^{{ab }}h_{{a(d}}^{{  }} t_{{c)b}}^{{  }}-
\frac{4}{3}t_{{cd}}\theta-t_{{cdae}}\sigma^{{ae}}-D_aQ^{a}_{\ cd},
\label{lie-t}\\
\fl\pounds_nQ_{{bcd}}^{{   }}=-A^a (t_{{bcda}}^{{    }}
+3h_{{a(d}}^{{  }}t_{{bc)}})+6
\omega^{{ae }}h_{{a(d}} Q_{{bc)e}}
-\theta_{}Q_{{bcd}}-D_at^{a}_{\ bcd}.
\label{lie-q}
\end{eqnarray}
\label{superenergylaw-vacuum}
\end{theo}
\P This can be regarded as a particular case of theorem 
\ref{superenergy-law} with ${\mathfrak J}_{abc}=0$ and 
$B_{abcd}=T_{abcd}$. This entails $t_{ab}=t^*_{ab}$, 
$\overline{Q}_{abc}=Q_{abc}$, $\overline{t}_{abcd}=t_{abcd}$,
$L_a=0$, $\overline{J}_{ab}=0$, $\tilde{J}_{ab}=0$, $j_{abc}=0$
which used in (\ref{tb-flow}) and (\ref{qb-flow}) 
leads to (\ref{lie-t}) and (\ref{lie-q}). Equation (\ref{tstar-constraint}) 
becomes an identity and (\ref{poynting-flow}) is now obtained by 
taking the trace of (\ref{tb-flow}). \N

In the particular case studied in theorem (\ref{superenergylaw-vacuum})
we find that (\ref{w-flow}) and (\ref{poynting-flow}) acquire simpler 
expressions which are
\begin{eqnarray}
\fl\pounds_nW=-4A^a{\cal P}_a-2W\theta-3t^{ae}\sigma_{ae}-D_a{\cal P}^a,
\label{cw-flow}\\
\fl\pounds_n{\cal P}_d=-A^a(3t_{da}+h_{da}W)-\frac{5\theta}{3}{\cal P}_d-
2Q_{dae}\sigma^{ae}+2{\cal P}^a\omega_{da}-D_at^a_{\ d}. 
\label{p-flow}
\end{eqnarray}
The linearized form of (\ref{cw-flow}) was known to Bel \cite{BEL4}
and in fact he took this equation as the starting point for a definition
of a state of {\em intrinsic radiation} for the gravitational field in vacuum
(see subsection 
\ref{bel-theory} for further details). The general form
of (\ref{cw-flow}) was derived in \cite{MAR-BAS}. 
It is interesting to note the formal analogy of 
(\ref{cw-flow})-(\ref{p-flow}) 
with (\ref{poynting-1})-(\ref{poynting-2}) where 
$W$ and ${\cal P}^a$ take the role of the electromagnetic energy density and
the Poynting vector respectively. Although  
(\ref{p-flow}) has, as far as we know,
never been obtained in its complete form, the knowledge of (\ref{cw-flow}), 
even in its linearized form, shown in equation (\ref{linear-form}), 
has been enough to construct the analogy
just mentioned and a lot of work has been devoted to studying the behaviour of 
gravitational systems by studying the super-energy density and the super
-Poynting vector in the system --see for example \cite{IBANEZ-VERDAGUER,BRETON,
WHEELER,HERRERA-POYNTING}. The results obtained are very suggestive 
but we must note that (\ref{cw-flow})-(\ref{p-flow}) are not 
equivalent to (\ref{lie-t})-(\ref{lie-q}) which in fact contain more 
information. Therefore, if we are to study gravitational radiation
 by means of techniques involving the study of the evolution of the different
spatial parts of Bel-Robinson tensor then we should start with the general
equations (\ref{lie-t})-(\ref{lie-q}). This matter is addressed in 
section \ref{superenergy_balance}.

\section{Application: superenergy radiative states of the gravitational field}
\label{superenergy_balance}
In electromagnetism, we speak of {\em electromagnetic radiation}
to mean that electromagnetic energy is traveling from one part of 
a system to another which in turn implies
the existence of a flux of energy-momentum.
By the Poynting theorem this flux is represented by 
the Poynting vector and thus whenever the Poynting vector is not 
zero at a point we say that electromagnetic radiation is going through
that point. This statement is observer dependent because in order to define
the Poynting vector an observer $n^a$ is needed (see equation 
(\ref{tem-decomposition})).
Therefore we may find for example, that the Poynting
vector is zero for one observer whereas another observer measures a 
non-vanishing Poynting vector. 
However, there are configurations in which any observer
will measure a non-vanishing Poynting vector and in these cases
it is said that the electromagnetic field is in a radiation state at the 
point. From an algebraic point of view this can only happen if the 
electromagnetic field $F_{ab}$ is {\em singular} or {\em null}
 which means that it can 
be written as the exterior product of a null and a spatial vector.    
(see e.g. \cite{NABER}).

If we try to follow the same procedure to define gravitational radiation
in General Relativity we are immediately confronted with the fact that, 
due to the equivalence principle, we can always find an observer who 
measures no ``gravitational energy density'' at a point, for any quantity 
with dimensions of energy constructed from the metric tensor $g_{ab}$
(typically this involves expressions which are quadratic in the first 
derivatives of the metric tensor). This 
means that in General Relativity we cannot pursue the same procedure used
to define {\em radiating fields} as in electromagnetism
{\bf if we insist upon using quantities
with dimensions of energy for this purpose}. Of course, this does not imply
that ``gravitational energy'' is meaningless and in fact we can construct
quasilocal and global quantities with dimensions of energy
 which tell us when a gravitational system is radiating. 
This has been performed for the important case of
 isolated systems where the quantity is the Bondi mass
\cite{BONDI,SACHS,PENROSE}.    

If instead of energy,
we use superenergy as a replacement, then the afore-mentioned 
problem disappears and one
can use the same ideas as in electromagnetism to define {\em radiating 
gravitational fields} or {\em radiating spacetimes} in a local 
way. This approach was 
pioneered by Bel many years ago in \cite{BEL4} and, indeed, 
the results presented
in this section can be regarded as a continuation of
Bel's work. We must bear in mind all the time that radiating gravitational
fields defined in terms of superenergy are in principle different from 
radiating fields defined by a quasilocal energy prescription. 
To find the precise relation between both concepts is 
an interesting open question which is a particular case of a more
general problem, namely, the possible relationship between superenergy and
energy. This is a long standing question which has been 
already largely researched 
\cite{BERGQVIST-LUDVIGSEN,BERGQVIST-SPHERE,HOROWITZ,MASHHOON-1,MASHHOON-2} 
(a fuller list of references about this subject can be found in 
\cite{SUPERENERGY}).

\subsection{Superenergy radiative states for vacuum spacetimes}
\label{bel-theory}
Let us start by reviewing Bel's work about the definition of 
a radiative spacetime.  
The starting point of Bel's study was the linearized form of 
(\ref{cw-flow}). To obtain this form, we define a coordinate chart
$(t,x^i)$, $i=1,2,3$ in such a way that $\partial/\partial t$ is 
the unit timelike vector $n^a$ and $\{\partial/\partial x^i\}$ 
are spacelike $\forall i$. 
Next we approximate the spatial covariant derivative 
by a covariant derivative compatible with the frame 
$\{\partial/\partial x^1,\partial/\partial x^2,\partial/\partial x^3\}$,
and ignore terms containing kinematical quantities. Under this 
approximation, equations (\ref{cw-flow})-(\ref{p-flow}) become
\begin{equation}
\frac{\partial W}{\partial t}
+\sum^3_{i=1}\frac{\partial{\cal P}^i}{\partial x^i}=0,\ 
\frac{\partial {\cal P}_i}{\partial t}-
\sum_{j=1}^3\frac{\partial t^j_{\ i}}{\partial x^j}=0.
\label{linear-form}
\end{equation} 
 These equations can always be obtained at a given point 
$p$ of the spacetime if we choose an observer $n^a$ such 
that all its kinematical quantities vanish at $p$ (such an
observer always exists according to the equivalence principle).
The first equation of 
(\ref{linear-form}) has the form of a typical conservation law. 
The vector ${\cal P}^i$ is, according to this equation, the flux of 
$W$ (superenergy flux) and whenever ${\cal P}^i$ is zero we see that 
$W$ does not change for the observer $\partial/\partial t$.
According to proposition \ref{weyl-zero} the superenergy density $W$ is
zero if and only if $C_{abcd}$ vanishes as well and besides $W$ is always 
nonnegative. Therefore, it is possible to take $W$ as a replacement for the 
missing concept of ``energy density'' of the gravitation and 
we may consider that
the existence of a flux of superenergy for any observer is an indication of
the intrinsic presence of gravitational radiation. These ideas led Bel to 
the following definition \cite{BEL4}.
\begin{defi}[State of intrinsic gravitational radiation, Bel 1962.]
We say \\ that  there is a state of intrinsic gravitational radiation at a 
point $p\in V$ of a vacuum spacetime
if ${\cal P}_a={\cal P}_a(n)$ does not vanish at $p$ 
for any $n^a$.
\label{bel-definition}
\end{defi}
A well-known consequence of definition \ref{bel-definition} is that 
Petrov types N, II and III are always radiative. To show this it is
enough to recall that the condition ${\cal P}^a=0$ entails (\ref{e-b})
which can only be true for either type I or type D.
Note that definition 
\ref{bel-definition} does not say anything about the radiative 
character of Petrov types I and D 
and in fact a more general definition would be needed to 
decide the issue. To obtain a generalization of definition 
\ref{bel-definition} is our next task.

To generalize definition \ref{bel-definition} we need to use the full 
information coming from the orthogonal splitting of $\nb_aT^a_{\ bcd}=0$
and not just (\ref{cw-flow}) which only contains part of this information.
Theorem \ref{superenergylaw-vacuum} contains all that is needed
in our endeavour. If we wish to use the variation of 
superenergy as a tool to define radiative states then we need 
to find the evolution of a 
spatial tensor whose vanishing is equivalent to the absence of a gravitational 
field (in vacuum this is just the condition $C_{abcd}=0$).
Bel's definition is based on the scalar $W$ but proposition 
(\ref{weyl-zero}) tells us that the tensor $t_{ab}$ plays
a similar role (and besides $W$ is not independent of $t_{ab}$).
The propagation of $t_{ab}$ is given by (\ref{lie-t}) and
we see that the only term in this equation not affected by kinematical
quantities (and hence intrinsic)
is $D_aQ^a_{\ bc}$.
\begin{defi}[Intrinsic superenergy radiative state in vacuum]
In a vacuum spacetime there exists 
 an intrinsic superenergy radiative state at a point 
$p\in V$ if $Q_{abc}(n)$ does not vanish at $p$ 
for any unit timelike normal $n^a$.  
\label{superenergy-radiation}
\end{defi}  
\begin{Remark}\em
We use the name {\em superenergy radiative state} instead of Bel's
original name of {\em radiative state} in order to stress
the fact that our definition is based on gravitational superenergy. 
\end{Remark}
Note that there are more tensors which have the relevant properties of 
$t_{ab}$ explained above and therefore we could use their propagation as the 
starting point for a definition of superenergy radiative state. 
The consequence of this is
that definition \ref{superenergy-radiation}
admits alternative but equivalent formulations. To see an example, 
consider the spatial tensor
\begin{equation}
W_{ab}\equiv E_{ac}E^c_{\ b}+B_{ac}B^c_{\ b}.
\label{wab}
\end{equation}
Clearly, $W^a_{\ a}=W$ and $W_{ab}=0\Longleftrightarrow C_{abcd}=0$.
Moreover, for any spatial vector $x^a$, $W_{ab}x^ax^b$ is non-negative
by inspection. We find
that in terms of $W_{ab}$ equation (\ref{lie-t}) takes the equivalent form
\begin{eqnarray}
\fl\pounds_n W_{cd}=A^a(-h_{cd}{\cal P}_a+\frac{1}{2}h_{ad}{\cal P}_c
+\frac{1}{2}h_{ac}{\cal P}_d+2S_{cda})-4\varepsilon_{ab(d}W_{c)}^{\ b}
\omega^a-\frac{4}{3}\theta W_{cd}+\nonumber\\
+\left(\frac{t_{cdab}}{2}+h_{ca}h_{db}W+3h_{cd}W_{ab}\right)\sigma^{ab}+
D_aS_{cd}^{\ \ a},
\label{wl-flow}
\end{eqnarray}
where 
\begin{equation}
S_{cda}\equiv 2 B_{b(d}E_{c)e}\varepsilon_a^{\ eb}.
\label{s-vector}
\end{equation}
In view of (\ref{wl-flow}) we deduce that definition
\ref{superenergy-radiation} can be formulated by replacing $Q_{abc}$
with $S_{abc}$. In fact from (\ref{s-vector}) and (\ref{br-parts}) 
we deduce
$$
S_{cda}=\frac{1}{2}(Q_{acd}-h_{cd}Q^{b}_{\ ba}),\ 
Q_{acd}=2(S_{cda}-h_{cd}S^{b}_{\ ba}),
$$
from which we conclude that both 
$S_{abc}$ and $Q_{abc}$ contain the same information and thus they should be 
deemed equivalent. We may expect that any reasonable
definition of a superenergy radiative state 
should be formulated in terms of a spatial tensor which is equivalent 
to $Q_{abc}$. Any such tensor can be regarded as the gravitational
equivalent of electromagnetism's Poynting vector. The 
tensor $S_{abc}$ seems to be the simplest choice and one may 
adopt it as the basic geometric object measuring ``superenergy
flux''.

Another interesting aspect of (\ref{lie-t})-(\ref{lie-q}) or
(\ref{wl-flow}), already pointed out in 
remark \ref{coupling}, is the fact that they are written in such a
way that the couplings of the kinematical quantities to the
different spatial parts of the decomposition of the Bel-Robinson tensor
are apparent.  In our present context these couplings could be interpreted
as the effect on the superenergy radiation due to the acceleration,
the expansion, the shear and the rotation. At this point it is instructive
to compare equation (\ref{lie-t}) (or its equivalent (\ref{wl-flow})) with
its electromagnetic counterpart which is (\ref{poynting-1}). In the 
electromagnetic case we realize that the vorticity has no effect whatsoever
on the radiation of electromagnetic energy whereas it certainly influences
the radiation of superenergy because $\omega^a$ (or equivalently $\omega_{ab}$)
appears explicitly in (\ref{lie-t}).

\subsection{Superenergy radiative states for general spacetimes}
Using the ideas explained in the previous section we 
can formulate a definition of an intrinsic superenergy radiative state
that is similar to definition 
\ref{superenergy-radiation} but valid for a general spacetime.
In this case we need to study the evolution of a spatial quantity 
which is zero if and only if the Riemann tensor vanishes. As stated
in proposition \ref{riemann-flat} the tensor $\overline{t}_{ab}$
has the required properties and hence the terms appearing in
the evolution equation of $\overline{t}_{ab}$ should enable us to define the
concept of an intrinsic radiative state. The evolution
equation sought is (\ref{tb-flow}) and hence the inspection of this equation
leads us to the following
\begin{defi}[Intrinsic superenergy radiative state in a general spacetime]
There exists an intrinsic
superenergy radiative state at a point $p\in V$ 
if for any unit timelike vector $n^a$ it is the case that 
$\overline{Q}_{abc}(n)$ does not vanish at $p$.  
\label{radiation-state}
\end{defi}
Similar considerations as in the case of definition \ref{superenergy-radiation}
apply here. 

\section{Conclusions and open issues}
In this work we have obtained the full orthogonal splitting of 
the Bel tensor and its covariant divergence and we have particularized 
it to the important case of vacuum spacetimes where the Bel tensor 
becomes the Bel-Robinson tensor. This gives rise to the
dynamical laws of superenergy. The concept of 
a superenergy radiative state has been introduced. The work just 
presented opens new research lines which we believe are worth exploring.
Perhaps one of the most interesting issues is a {\em global} formulation
of the dynamical laws of superenergy complementing the local formulation of 
theorem \ref{superenergy-law}. Such a global formulation would enable us
to apply our techniques to realistic astrophysical settings such as oscillating
stars, rotating bodies or radiating binary systems. 

In this paper we have restricted ourselves to the superenergy
defined from the Riemann and Weyl tensor
but one can define tensors representing superenergy from a 
general field resulting in the {\em superenergy tensor} of
that field \cite{SUPERENERGY}. In this framework it is possible to
calculate the covariant divergence of a superenergy tensor and obtain 
an expression similar to the first equation in 
(\ref{mattercurrent-conservation}) with the Bel tensor replaced by 
a suitable superenergy tensor. The orthogonal splitting
of such an equation would yield the dynamical
laws of the superenergy associated with that particular field. 
An interesting example concerns the electromagnetic field.
In this case a possible superenergy tensor is the
Chevreton tensor which was first introduced in \cite{CHEVRETON}
and recently stimulating results about its symmetries and
the covariant divergence of its trace have been obtained \cite{BES}. 
The Chevreton tensor, like the Bel-Robinson
tensor, is a rank-four tensor and its covariant divergence couples
the Weyl tensor with terms which contain covariant derivatives of
the Faraday tensor \cite{INGEMAR}. This suggests a possible exchange between the
gravitational and the electromagnetic superenergies 
\cite{SUPERENERGY,LZV,ERIKSSON}. The orthogonal splitting of the 
covariant divergence of the Chevreton tensor might shed light on the nature
of this exchange.      

Another important issue is the possible relationship between superenergy
and any of the available quasilocal concepts of {\em gravitational energy} 
which have been developed over the years. 
This is a topic which has been extensively
researched in the past and no clear conclusion has been reached. In 
this work no attempt has been made in this direction and our point of view has 
been to regard superenergy as a physical quantity on its own right. We
believe that this idea can be put to work by means of the results of theorem
\ref{superenergy-law} which would demand a formulation of the 
dynamical laws of superenergy tailored for each physical system under study.
However, a relation between superenergy and gravitational energy cannot 
be ruled out and the orthogonal
splitting of the Bel tensor might bring a new point of view to this old 
problem. 

\section{Acknowledgements}
We wish to thank Jos\'e M M Senovilla for inspiring discussions
and his careful reading of the manuscript. We also thank Lluis Bel,  
Ingemar Eriksson  and Jos\'e M Mart\'\i n-Garc\'\i a for valuable comments.
Jos\'e M Mart\'\i n-Garc\'\i a is also thanked for his technical aid
with the system {\em xAct}.
Finally we thank the constructive criticism of two anonymous referees
which helped to improve a previous version of the manuscript.  
Financial support of the Spanish ``Ministerio de Educaci\'on y 
Ciencia'' under the postdoctoral fellowship EX2006-0092 is gratefully
acknowledged.  

\appendix
\section{Technical details about the computations}
In this appendix we supply details about the calculations 
required in this work. In order to do so we need to explain 
some implementation aspects of the system {\em xAct}. We will limit ourselves
to only those issues which are needed in our calculations referring the
interested reader to \cite{JMM} for a full documentation and tutorials
about {\em xAct}. 
  
Orthogonal splittings play an essential part in our work and the 
implementation of {\em xAct} in regard to this matter is completely adapted to
our requirements. The basic elements of the orthogonal splitting are
defined through the command
\small
\begin{verbatim}
In[]:= DefMetric[1, h[-a, -b], cd, {"|","D"}, InducedFrom->{g, n}, PrintAs->"h"].
\end{verbatim}
\normalsize
Here \verb|h[-a,-b]| represents the spatial metric $h_{ab}$
which is constructed from the spacetime metric $g_{ab}$ (represented in
the system by \verb|g[-a,-b]|) 
and the unit normal vector $n^a$ (represented by \verb|n[a]|).  
The operator \verb|cd[-a]| is the Cattaneo operator $D_a$ associated with 
$h_{ab}$. The system is able to handle all the properties of 
the Cattaneo operator explained in subsection \ref{section-cattaneo}
in a natural fashion.

The general expression for 
the orthogonal splitting of any tensor is equation (\ref{general-ot}).
This result is implemented in {\em xAct} by means of the command
\begin{verbatim}
In[]:= InducedDecomposition[expr, {h,n}],
\end{verbatim}
where \verb|expr| represents any tensorial expression.
The output of \verb|InducedDecomposition| is the result of applying 
formula (\ref{general-ot}) to \verb|expr|. 
The orthogonal projector operator $P_h$
which appears in (\ref{general-ot}) is also implemented in {\em xAct}
by means of the command \verb|Projectorh[expr]| where 
again \verb|expr| represents an arbitrary tensor. The
basic commands just explained enable us to find efficiently orthogonal
splittings similar to equation (\ref{l-split}) with 
$L_a$ replaced by any spatial tensor of higher rank. 

\subsection*{Proof of theorems \ref{superenergy-law} and \ref{matter-law}}
To prove theorems \ref{superenergy-law} and \ref{matter-law}  
we need to find the orthogonal decomposition of 
the equations shown in (\ref{mattercurrent-conservation}). 
The first step is to replace 
$B_{abcd}$, $R_{abcd}$ and ${\mathfrak J}_{abc}$ 
with their orthogonal splittings, eqs. (\ref{beltensor-split}), 
(\ref{riemann-split}) and (\ref{matter-current-decomposition}) respectively.
The covariant derivatives of $n^a$ are decomposed according to 
(\ref{decompose-normal}) and $\nb_a\varepsilon_{bcd}$ is decomposed 
by means of the formula
\begin{eqnarray}
\nb_d\varepsilon_{fhl}=3A^an_dn_{[h}\varepsilon_{fl]a}-3
n_{[h}\left(\frac{1}{3}\varepsilon_{fl]d}\theta+
\varepsilon_{fl]a}(\sigma_d^{\ a}+\omega_d^{\ a})\right).
\nonumber
\end{eqnarray}
After doing these replacements we obtain expressions which contain
$\nb_a\overline{W}$, $\nb_a\overline{\cal P}_b$, $\nb_a\overline{t}_{bc}$, 
$\nb_at^*_{bc}$, $\nb_a\overline{t}_{bcde}$, 
$\nb_aL_b$, $\nb_a\tilde{J}_{bc}$, $\nb_a\overline{J}_{bc}$, $\nb_aj_{bcd}$.
These are further decomposed by following the procedure explained in 
subsection \ref{section-cattaneo}. For example, the orthogonal decomposition
of $\nb_aL_b$ is just equation (\ref{l-split}) which also holds if we 
replace $L_b$ with $\overline{\cal P}_b$. Other
orthogonal decompositions needed are 
\begin{eqnarray}
\fl\nb_c\overline{t}_{ab}=-2A^dn_cn_{(b}\overline{t}_{a)d}+
2n_{(a}\left(\frac{1}{3}\overline{t}_{b)c}\theta+\overline{t}_{b)}^{\ d}
(\sigma_{cd}+\omega_{cd})\right)+D_c\overline{t}_{ab}+\nonumber\\
+n_c\left(\frac{2}{3}\theta\overline{t}_{ab}+
2\overline{t}_{(b}^{\ d}(\sigma_{a)d}+\omega_{a)d})
-\pounds_n\overline{t}_{ab}\right),\label{oth-t}
\end{eqnarray}
which is also valid if we replace $\overline{t}_{ab}$ by any symmetric 
spatial tensor and
\begin{eqnarray}
\fl\nb_a\tilde{J}_{bc}=2A^dn_a(\tilde{J}_{d[b}n_{c]})+
2n_{[c}\left(\frac{1}{3}\tilde{J}_{b]a}\theta+\tilde{J}_{b]}^{\ d}
(\sigma_{ad}+\omega_{ad})\right)+D_a\tilde{J}_{bc}+\nonumber\\
+n_a\left(\frac{2}{3}\tilde{J}_{bc}\theta
+2\tilde{J}_{[b}^{\ c}(\sigma_{c]d}+\omega_{c]d})-\pounds_n\tilde{J}_{bc}
\right),
\end{eqnarray}
which is true if we replace $\tilde{J}_{bc}$ with any antisymmetric tensor.
The expressions for the orthogonal splitting of the remaining covariant 
derivatives are very long and we omit them. Inserting the orthogonal splittings
in (\ref{mattercurrent-conservation}) and rearranging the equations obtained 
as polynomials in $n^a$ we obtain the expressions
\begin{eqnarray}
\fl An^bn^cn^d+B_1^{b}n^cn^d+B_2^{(c}n^{d)}n^b+
C_1^{cd}n^b+C_2^{b(c}n^{d)}+
E^{bcd}=0,\label{polinomial-1}\\
G_1^{[b}n^{c]}+I^{bc}=0,
\label{polinomial-2}
\end{eqnarray}
where all the tensor coefficients of these polynomials are spatial. 
This implies that the coefficients of the polynomials
must vanish and these conditions lead us to 
\begin{equation}
\fl A=0,\ B_1^{b}=B_2^{b}=0,\ C_1^{(cd)}=C_1^{cd}=0,\ C_2^{bc}=0,\ 
E^{bcd}=0,\ G_1^b=0,\ I^{[bc]}=I^{bc}=0.
\label{pol-zero}
\end{equation}
After some manipulations we find that the condition $A=0$ is equivalent
to (\ref{w-flow}), $B_1^{b}=B_2^{b}=0$ are equivalent to 
(\ref{poynting-flow})-(\ref{tstar-constraint}), $C_1^{cd}=0$ is 
equivalent to (\ref{tb-flow}), $C_2^{bc}=0$ is equivalent to (\ref{ts-flow}),
$E^{bcd}=0$ is equivalent to (\ref{qb-flow}), $G_1^b=0$ is equivalent to
(\ref{lb-flow}) and $I^{bc}=0$ is equivalent to (\ref{jtilde-flow}). 
The condition $A=0$ is redundant because it can be obtained as the 
trace of $C_1^{cd}=0$ and therefore we do not need to consider it. We have 
thus recovered all the expressions given in the statements of
theorems \ref{superenergy-law} and \ref{matter-law}. Note that 
the polynomials (\ref{polinomial-1})-(\ref{polinomial-2}) are equivalent
to each of the equations presented in (\ref{mattercurrent-conservation})
and so is the set of conditions stemming from (\ref{pol-zero}).\N

\section{Canonical forms for the electric and 
magnetic parts of Weyl tensor in the different Petrov types.}
We present next the canonical forms of the electric and 
magnetic parts of Weyl tensor for the different Petrov types. 
We follow \cite{BEL4} in our presentation (see also 
\cite{MCCALLUM} for an equivalent representation of the 
canonical forms). All the canonical forms 
are written with respect to a certain orthonormal frame 
$O\equiv\{e^a_1,e^a_2,e^a_3\}$ of spatial vectors (canonical frame). 

\subsection*{Petrov type I}
In this type $E_{ab}$ and $B_{ab}$ take the following 
form in the canonical frame $O$ 
\begin{equation}
E_{\bf ab}=\mbox{diag}(E_{11},E_{22},E_{33}),\ 
B_{\bf ab}=\mbox{diag}(B_{11},B_{22},B_{33}),
\label{type-I}
\end{equation}
with the additional conditions 
\begin{equation}
E_{11}+E_{22}+E_{33}=0,\ B_{11}+B_{22}+B_{33}=0.
\end{equation}
\subsection*{Petrov type D}
This type arises if we set $-\frac{1}{2}E_{11}=E_{22}=E_{33}$,
$-\frac{1}{2}B_{11}=B_{22}=B_{33}$ 
in the previous case.
\subsection*{Petrov type II}
The canonical forms for $E_{ab}$, $B_{ab}$ in the frame $O$ are
\begin{eqnarray}
\fl E_{\bf ab}=
\left(\begin{array}{ccc}
                            E_{11} & 0 & 0\\
     0       &-\frac{E_{11}}{2}+B_{23} & E_{23}\\
0   & E_{23} & -\frac{E_{11}}{2}-B_{23}        
\end{array}\right),\nonumber\\ 
\fl B_{\bf ab}=
\left(\begin{array}{ccc}
                            B_{11} & 0 & 0\\
     0       &-\frac{B_{11}}{2}-E_{23} & B_{23}\\
0   & B_{23} & -\frac{B_{11}}{2}+E_{23}        
\end{array}\right).
\label{type-II}
\end{eqnarray}
\subsection*{Petrov type III} 
The canonical forms for $E_{ab}$, $B_{ab}$ in the frame $O$ are 
\begin{equation}
\fl E_{\bf ab}=
\left(\begin{array}{ccc}
                            0 & E_{12} & -B_{12}\\
     E_{12}       &0 & 0\\
-B_{12}   & 0 & 0       
\end{array}\right),\ 
B_{\bf ab}=
\left(\begin{array}{ccc}
                            0 &B_{12}& E_{12}\\
     B_{12}   &0 & 0\\
E_{12} &0 & 0       
\end{array}\right).
\label{type-III}
\end{equation}
\subsection*{Petrov type N} 
The canonical forms for $E_{ab}$, $B_{ab}$ in the frame $O$ are 
\begin{equation}
\fl E_{\bf ab}=
\left(\begin{array}{ccc}
                            0 & 0 & 0\\
     0       &E_{22} & -B_{22}\\
0   & -B_{22} & -E_{22}       
\end{array}\right),\ 
B_{\bf ab}=
\left(\begin{array}{ccc}
                            0 & 0& 0\\
     0   &B_{22} & E_{22}\\
0 &E_{22} & -B_{22}       
\end{array}\right).
\label{type-N}
\end{equation}

\end{document}